\documentclass[journal]{IEEEtran}
\usepackage{amsmath,amssymb,graphicx,booktabs,array,url}
\usepackage[caption=false,font=footnotesize]{subfig}

\newcommand{\GF}{\mathrm{GF}(2)}
\newcommand{\T}{\mathsf{T}}

\begin{document}
\title{Protocol Data Unit Header Decoding and Payload Coding Recovery in the Starlink Ku-Band Downlink}
\author{Haoyan Liu, Yichuan Ma, Haotong Zhao, Chengguang Li, and Min Yang%
\thanks{The authors are with the School of Aerospace Science and Technology, Xidian University, Xi'an 710071, China.  (e-mail: haoyan.xa@gmail.com; 23131110622@stu.xidian.edu.cn; 25131213466@stu.xidian.edu.cn; chgli@stu.xidian.edu.cn; merovingia1911@126.com.)}
}\maketitle
\begin{abstract}
Existing Starlink downlink studies describe synchronization, pilots, and predictable waveform elements, but these descriptions do not establish how received header bits specify payload allocations and channel coding. Using 18 frames from one continuous complex-baseband capture spanning approximately 24 ms, we identify two protocol data unit (PDU) header regions in each frame and decode their contents after receiver correction and reference-phase removal. We recover the headers' terminated rate-1/3 convolutional code with constraint length seven and an eight-bit cyclic redundancy check over the concatenated information words. The decoded bits support a conditional 20-bit layout of modulation and coding scheme (MCS) index--length pairs, linking format labels and quantities to subsequent payload regions. Predicted allocation endpoints agree with signal-based estimates in 16 of 18 frames and differ by at most two symbols in the remaining frames. We further recover modulation mappings and equivalent quasi-cyclic low-density parity-check coding relations for several payload formats, including rate-1/2 and rate-5/8 families. Fixed generators predict 78,848, 9,984, and 811,008 parity bits for formats 5, 9, and 21 without disagreement on codewords excluded from their generator fitting. Cross-modulation tests support shared encoders. Formats 28 and 37 have verified shortened cores, with 256 and 384 additional bits unexplained, while format 68 remains unresolved.
\end{abstract}

\section{Introduction}
\IEEEPARstart{B}{roadband} low-Earth-orbit (LEO) transmissions provide opportunities for passive measurement beyond their intended communication service. Their timing, frequency, bandwidth, and repeatable waveform components can support a receiver that has no access to the transmitted application data. Doppler-based navigation concepts illustrate one motivation for studying such signals~\cite{psiaki2021}; detailed waveform knowledge also supports synchronization, interference characterization, and receiver reconstruction. These applications need different levels of knowledge. Knowing where an OFDM symbol begins is useful for acquisition, but it does not establish where a codeword begins, which carriers carry control information, or whether the remaining symbols obey an identified channel code. Moving from a physical waveform description to a reproducible description of its coded fields is therefore a separate identification problem.

Existing investigations of Starlink signals have examined synchronization, resource organization, pilot placement, and modulation and coding.  Jiao Xianjun's public technical analyses use uplink measurements to investigate repeated training intervals, OFDM boundaries, and candidate pilot subbands~\cite{jiao2026uplink}. His patent-based analysis also summarizes convolutionally encoded packet headers, LDPC payload coding, and pseudonoise phase rotation~\cite{jiao2026coding}. These reports provide useful context, but uplink observations and patent descriptions do not directly establish the configuration of a captured downlink frame. The present work therefore focuses on the satellite-to-terminal Ku-band downlink and its experimentally observed field organization.

Humphreys \emph{et al.}~\cite{humphreys2023} established the principal time–frequency parameters of the Starlink Ku-band downlink and disclosed its primary and secondary synchronization sequences. Their results provide the frame periodicity, OFDM numerology, and synchronization references needed to align received frames, while leaving finer logical subframe organization unresolved. Qin \emph{et al.}~\cite{qin2026} subsequently recovered edge-pilot sequences, a constant reference template, and predictable patterns within the data-bearing region. These findings support acquisition and demodulation with compact receivers and reveal regularity beyond the synchronization intervals. Together, they provide the physical foundation for examining how the remaining resources are divided into headers and coded payloads.

These studies establish a detailed physical description of the Starlink downlink, but a gap remains between recognizing its waveform and interpreting its coded contents. Public patents describe mechanisms such as convolutional header coding, payload LDPC encoding, and modulation and coding scheme (MCS) signaling~\cite{mccormick2024ofdm,iyer2023amc}. Such descriptions provide useful hypotheses, but do not establish how those mechanisms appear in a particular received stream. A waveform compatible with the published numerology does not, by itself, reveal the boundaries of its logical fields, the meaning of their bits, or the matrices defining its payload codes. The remaining task is to connect candidate field interpretations to measurable resource allocations and explicit coding relations, and to verify that these relationships consistently explain the received observations.

This task involves several coupled ambiguities. Predictable phase rotations and receiver impairments affect the apparent constellation and its symbol-to-bit mapping~\cite{dobre2007}. Even with a plausible mapping, an incorrect codeword boundary or interleaving assumption can obscure the redundancy of the underlying code. Conversely, a sufficiently flexible model can fit a limited collection of bit sequences without identifying their actual organization. Blind linear-code recognition~\cite{valembois2001,cluzeau2006}, convolutional-encoder identification~\cite{marazin2011}, and interleaver estimation~\cite{sicot2009} address different aspects of this problem. Their relevance here is that field extraction and coding reconstruction must be assessed jointly: candidate boundaries determine which bits enter a coding test, while the recovered coding relations provide evidence for or against those boundaries.

Finite sample diversity further limits what can be inferred. When observed inputs do not span the candidate information dimension, multiple generators can reproduce the same codewords. Structural assumptions can reduce this ambiguity, but agreement within the fitted model must be distinguished from evidence obtained outside its fitting equations. For LDPC identification, a successful generator fit must also be supported by a complete sparse parity-check representation~\cite{gallager1962,mackay1999}. QC structure offers a compact model for this recovery~\cite{fossorier2004,lally2001}, although cyclic shifts constructed under that assumption are not independent received transmissions. We therefore distinguish parameter fitting, structural-model selection, and subsequent verification, with the data used at each stage stated explicitly~\cite{cawley2010}.

Our analysis proceeds from observable signal organization to progressively more specific bit-level interpretations. After synchronization, equalization, and reference-phase removal, we compare modulation patterns and recurring resource lengths across frames. These observations suggest candidate binary intervals and block sizes, which are tested through convolutional relations, state termination, and re-encoding. CRC consistency then establishes whether adjacent decoded blocks form a common checked unit. Only after these relationships have been verified do we examine the decoded fields and their correspondence with subsequent modulation regions and resource lengths. This correspondence supports interpreting two recovered units as protocol data unit (PDU) headers. Their candidate MCS fields then constrain payload extraction, while parity prediction and remodulation test whether the proposed allocations and coding relations explain the same received signal.

The contributions are threefold. First, we recover two CRC-checked units in each frame, establish their terminated rate-1/3 convolutional encoding with constraint length seven, and identify their role as PDU headers through their relationship with subsequent resource allocations. Second, we obtain a conditional MCS index–length layout that links decoded header bits to observed modulation and symbol counts. The accumulated lengths agree with signal-estimated payload endpoints in 16 of 18 frames and differ by one or two symbols in the remaining frames. Third, we recover equivalent payload-code representations and validate them through fixed-generator parity prediction, cross-modulation transfer, re-encoding, and remodulation. For three formats, codewords excluded from generator fitting yield zero disagreement over 78,848, 9,984, and 811,008 predicted parity bits, respectively. These results connect signal structure, decoded fields, and channel coding within one consistent resource description.

The paper follows the identification dependencies. Section II reviews the frame structure and receiver corrections, including PN reference removal. Section III recovers terminated coding blocks and their group CRC. Section IV interprets control fields and tests resource allocations. Section V recovers equivalent QC encoders and validates shared encoders across modulation formats, including shortened cores. Section VI compares the results with published patent configurations. Appendix~\ref{app:regions} provides complete region maps and reference-removed constellations for all 18 frames. This is a reasoning order, not an invented chronological experiment log.

\section{Frame-Structure Review and Received-Signal Equalization}
\label{sec:receiver}
\subsection{Signal-Format Review}

\begin{figure*}[t]\centering
\subfloat[]{\includegraphics[width=.94\textwidth]{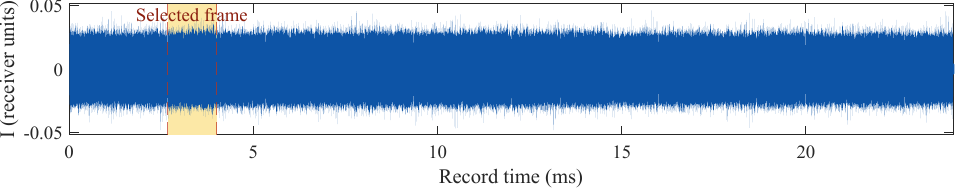}}\\[-2pt]
\subfloat[]{\includegraphics[width=.94\textwidth]{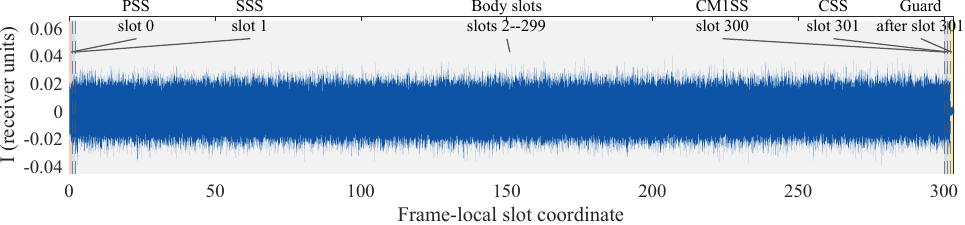}}\\[-2pt]
\subfloat[]{\includegraphics[width=.44\textwidth]{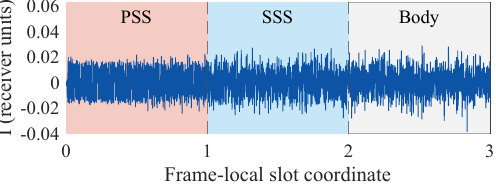}}\hfill
\subfloat[]{\includegraphics[width=.44\textwidth]{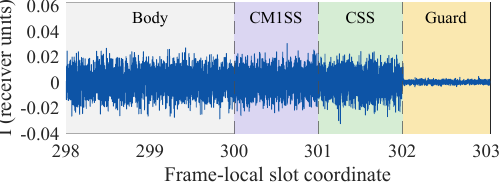}}
\caption{Temporal organization of the received signal. (a) The entire received baseband waveform, with R02 highlighted. (b) The complete R02 waveform and its temporal organization. (c) Frame acquisition region: PSS, SSS, and the start of the body. (d) Frame termination region: final body slots, published CM1SS/CSS positions, and the following guard. Envelopes in (a)--(b) preserve extrema in 256- and 32-sample bins; (c)--(d) show original in-phase samples. CM1SS/CSS denote published positions without assuming invariant terminal content in every observed frame.}\label{fig:recordtime}
\end{figure*}

The analyzed signal is one continuous complex-baseband capture after downconversion and upstream coarse carrier synchronization. With $x[n]=I[n]+jQ[n]$, the stated sampling rate is $F_r=480$ MHz; 11,520,180 samples span 24.000375 ms and contain 18 analyzed complete frames. Storage partitioning does not define separate signals, acquisition conditions, or validation datasets. Throughout the paper, R02 is the principal worked example.

The physical-layer structure we adopt mostly follows Humphreys \emph{et al.}~\cite{humphreys2023}, together with the edge-pilot coefficients reported by Qin \emph{et al.}~\cite{qin2026}. As shown in Fig.~\ref{fig:recordtime}, a nominal 750 Hz frame comprises 302 occupied slot intervals: the special time-domain PSS in slot 0, the OFDM SSS in slot 1, and a further 300 OFDM symbols in slots 2--301. At our receive rate, each ordinary symbol is processed with a 2048-point FFT and a 64-sample cyclic prefix, which is equivalent to 1024/32 at 240MHz. The subcarrier spacing is 234.375 kHz, and each slot lasts $4.4~\mu s$. A nominal frame thus leaves a $4.5333\mu s$ guard interval between frames. Humphreys \emph{et al.} observed that, some subcarrier symbols repeated across frames while the remaining symbols varied in slot 300, and referred to this predictable subset as the coda-minus-one synchronization sequence (CM1SS). Their study did not specify the complete carrier locations or symbol values of this subset. Qin \emph{et al.} later recovered the explicit sequences of 16 edge pilots, located on eight subcarriers near each edge of the occupied band and transmitted throughout slots 2--301. The edge pilots in slot 300 may therefore account for part of the predictable content previously associated with the CM1SS, although the available evidence does not establish that the CM1SS consists exclusively of these pilots. Humphreys \emph{et al.} further identified slot 301 as the coda synchronization sequence (CSS), which appeared as a known QPSK OFDM symbol in their capture and was followed by the inter-frame guard~\cite{humphreys2023,qin2026}. Our capture follows the same slot grid, but the non-pilot symbols in its terminal slots are not invariant across all frames.

Symbol loaded carriers are $\mathcal K=\{-512,\ldots,-3\}\cup\{2,\ldots,511\}$, with four central nulls and edge pilots at $\mathcal K_p=\{-496,\ldots,-489\}\cup\{488,\ldots,495\}$. The 16 pilots leave 1004 non-pilot modulation symbols per data slot. Later fields use $(s,q,k)$: slot $s$, non-pilot carrier ordinal $q$, and symbol carrier $k$. The equivalent ordered coordinate is
\begin{equation}\label{eq:coordinate}
r=1004(s-2)+q.
\end{equation}
The following sections identify where each control and payload field begins and ends, locate regions filled with the known reference sequence, and determine codeword boundaries and systematic/parity-bit layouts.

\begin{figure}[t]\centering
\subfloat[]{\includegraphics[width=\columnwidth,trim={17bp 7.5bp 0bp 0bp},clip]{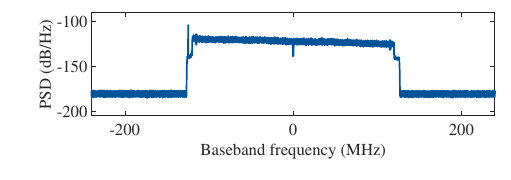}}\\[2pt]
\subfloat[]{\includegraphics[width=\columnwidth,trim={17bp 7.5bp 0bp 0bp},clip]{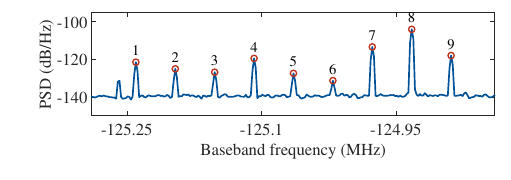}}
\caption{Welch PSD of the provided signal. (a) Full-band received-signal spectrum. (b) Enlargement of the nine-tone reference region outside the loaded OFDM grid. Numbered peaks identify the regularly spaced comb; the nearby unnumbered peak is outside that comb.}\label{fig:recordspectrum}
\end{figure}

By using Welch's method~\cite{welch1967}, full-band received-signal spectrum and enlargement of the nine-tone reference region outside the loaded OFDM grid are shown in Fig.~\ref{fig:recordspectrum}. A 262144-sample Hann window with 50\% overlap gives a frequency-bin spacing of 1831.055 Hz. The enlarged spectrum reveals nine narrowband tones near $-125$ MHz, separated by approximately 43.945 kHz. These tones lie outside the loaded OFDM band and are separate from the 16 modulated edge-pilot subcarriers. We track their frequency variation to check the frequency-offset estimates obtained from the OFDM pilots. However, their nominal transmitted frequencies are unknown, so the tones alone cannot determine the absolute Doppler shift. Numbered peaks identify the regularly spaced comb; the nearby unnumbered peak is outside that comb. PSD is relative to receiver-unit squared per hertz, not calibrated RF power.

\subsection{Received-Signal Equalization}
After frame acquisition, we describe the provided baseband signal by
\begin{equation}\label{eq:receiver}
r(t)=e^{j(2\pi f_D\tau+\pi\dot f_D\tau^2)}(h*x)(t-\epsilon(t))+w(t),
\end{equation}
where $j=\sqrt{-1}$, $*$ denotes convolution, and $t_0$ is the capture midpoint. The signal $x(t)$ is the transmitted waveform, $h$ is the effective channel response, and $w(t)$ is additive noise. With $\tau=t-t_0$, $f_D$ is the frequency offset at the capture midpoint and $\dot f_D$ is its rate of change. The timing error $\epsilon(t)$ includes an initial offset and sampling drift. This model approximates the frequency evolution over the 24ms record by a linear trend. We estimate the Doppler compensation from the CP and pilot observations below. The estimated offset may also include oscillator and transmitter errors, which this capture does not separate from propagation Doppler.

\begin{figure}[t]\centering
\subfloat[]{\includegraphics[width=.48\columnwidth]{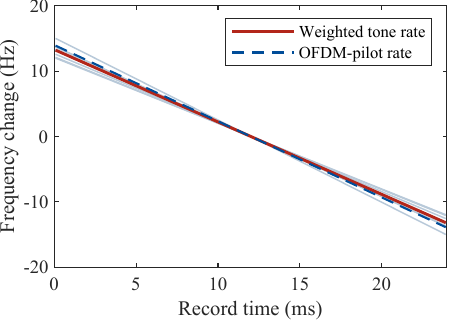}}\hfill
\subfloat[]{\includegraphics[width=.48\columnwidth]{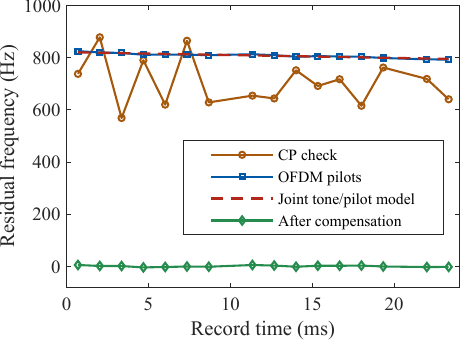}}
\caption{Tone-assisted receiver correction. (a) Baseline-removed frequency changes from eight accepted tracks in the nine-tone bank (light curves), their power-weighted trend, and the OFDM-pilot rate comparison. The weak sixth tone is excluded. Unknown tone baselines do not provide absolute Doppler measurements. (b) CP check, known OFDM-pilot estimates, the model combining tone rate with pilot-anchored offset, and actual post-compensation residual (RMS 2.742306 Hz).}\label{fig:receivercorrection}
\end{figure}

We first obtain a coarse frequency estimate from CP repetition~\cite{vandebeek1997,schmidl1997}. A CP sample and its repeated counterpart are separated by one useful-symbol duration $T_u$, so their correlation phase measures the carrier rotation over that interval:
\begin{equation}\label{eq:cpfreq}
\widehat f_{\rm CP}=\frac{\arg\sum_{t\in\mathcal I}r(t+T_u)r(t)^*}{2\pi T_u}.
\end{equation}
The set $\mathcal I$ contains the selected CP positions, $\arg$ returns the principal complex phase, and the superscript $*$ denotes complex conjugation. For initial acquisition, we form the products of each sample and the complex conjugate of the sample 2048 positions earlier on the preliminary resampled grid. We then average these products at corresponding positions across 2112 sample OFDM slots. Summing the 35 positions with the largest correlation magnitudes gives $\widehat f_c=47.449347$ kHz. We retain the remaining frequency error for pilot refinement. A separate CP check uses the last 32 CP samples of ordinary OFDM symbols on the refined receive grid. After the initial correction, its median estimate is 705.056 Hz as shown in Fig.~\ref{fig:receivercorrection}(b). The two CP measurements use different sample selections and serve as coarse estimates.

The nine narrow tones in Fig.~\ref{fig:recordspectrum}(b) provide the frequency-rate estimate. Each tone is mixed to its measured coarse center and tracked with 65536-sample Hann windows at 16384-sample intervals. A quadratic fit to its unwrapped phase gives
\begin{equation}\label{eq:tonerate}
\theta_i(\tau)\simeq a_i+b_i\tau+c_i\tau^2,\qquad
\widehat{\dot f}_D=\frac{1}{\pi}\sum_{i\in\mathcal T}w_i c_i.
\end{equation}
Here $i$ indexes a tone, $\theta_i$ is its unwrapped phase, $a_i,b_i,c_i$ are the fitted polynomial coefficients, $\mathcal T$ is the set of retained tracks, and $w_i$ is a normalized power weight. We retain tracks with phase-fit RMS below 0.5 rad and weight them by their mean powers, with $\sum_i w_i=1$. Eight of the nine tracks pass this test. The weak sixth tone is excluded because of phase cycle errors. The resulting rate is $-1107.157$ Hz/s. The nominal transmitted tone frequencies are unknown, so the fitted linear phase terms alone cannot establish the common frequency offset.

Next, we estimate that offset from the 16 known OFDM edge pilots. After removing their known coefficients and the static channel response, we fit their phase evolution over slots 17--289. Fitting the carrier-wise frequency estimates against subcarrier index separates the common offset from sampling-related phase drift. At the capture midpoint, the remaining offset is $\widehat f_r=808.639$ Hz. The OFDM pilots independently provide a rate estimate of $-1165.205$ Hz/s for comparison with the tone-derived rate in Fig.~\ref{fig:receivercorrection}(a). Both estimates use the same capture.

Adding the pilot refinement to the CP estimate gives the frequency correction applied to the provided IQ:
\begin{equation}\label{eq:dopplercompensation}
\begin{aligned}
\widehat f_D(\tau)&=\widehat f_c+\widehat f_r+\widehat{\dot f}_D\tau\\
&\simeq 48257.986-1107.157\tau\quad\text{Hz},\\
r_c(t)&=r(t)e^{-j[2\pi(\widehat f_c+\widehat f_r)\tau+\pi\widehat{\dot f}_D\tau^2]}.
\end{aligned}
\end{equation}
Here $\tau$ is in seconds. In implementation, the coarse correction is applied first and the residual chirp correction follows. Referencing both to $t_0$ changes only a constant phase. Fig.~\ref{fig:receivercorrection}(b) compares the estimated frequency trend with the CP and OFDM-pilot measurements. Repeating the FFT and pilot frequency measurement after compensation gives a residual RMS of 2.742306 Hz. This is a check of the applied correction on the estimation capture, not an independent Doppler calibration.

We then correct timing drift by resampling each frame on its estimated fractional sampling grid. The relative timing RMS in the illustrated frame decreases from 12.178511 to 0.052661 ns. After CP removal and FFT, the SSS provides the channel estimate used for equalization:
\begin{equation}\label{eq:equalization}
\begin{aligned}
\widehat H_k&=\mathcal S\{Y_{1,k}/S_k\},\\
\widehat X_{s,k}&=\frac{Y_{s,k}}{\widehat H_k}e^{-j(\phi_s+\alpha_s k)}.
\end{aligned}
\end{equation}
Here $Y_{s,k}$ is the corrected FFT output and $S_k$ is the known SSS coefficient. The smoothing operator $\mathcal S$ exploits the effective channel's frequency correlation~\cite{edf1998}. The edge pilots estimate the remaining common phase $\phi_s$ and linear phase slope $\alpha_s$ in each slot.

\begin{figure}[t]\centering
\subfloat[]{\includegraphics[width=.48\columnwidth]{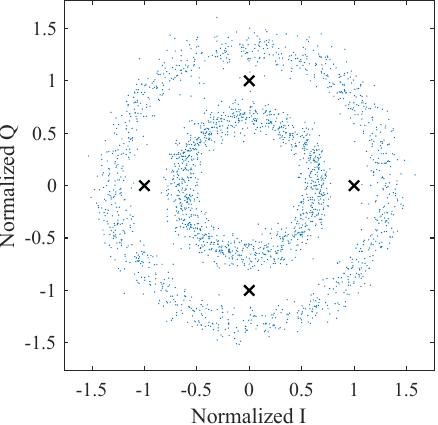}}\hfill
\subfloat[]{\includegraphics[width=.48\columnwidth]{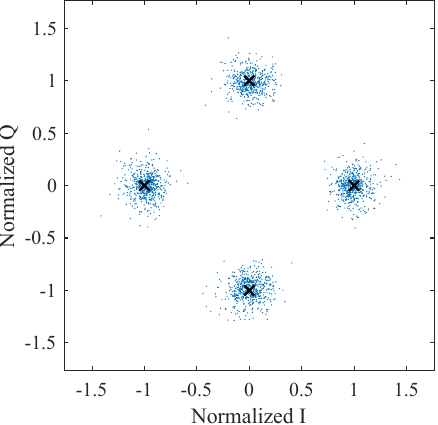}}
\caption{Matched pilot observations over slots 17--289: (a) coarse-corrected FFT values with scalar amplitude normalization and (b) fully equalized measurements.}\label{fig:equalizedpilots}
\end{figure}

Fig.~\ref{fig:equalizedpilots} shows the received pilot constellations before and after equalization. The eight pilot carriers excluded from the per-slot phase fit have a median EVM of 0.154289. A phase trend fitted over slots 50--289 also predicts the early pilot labels in slots 2--16 with 0 errors among 3840 observations. These checks establish the receiver output used for the subsequent analysis. Position-dependent data-reference removal is described in Section II-C.

\subsection{PN Phase Rotation}
\label{sec:reference}
SpaceX patent US12074683B1 describes a frequency-domain scrambler that rotates channel-estimation, payload, and pilot symbols by multiples of $\pi/2$ using a 15-stage shift register~\cite{mccormick2024ofdm}. Qin \emph{et al.} report predictable pilot coefficients and a reference template that exposes binary variations in some four-point regions~\cite{qin2026}. We test a compatible reference against this capture to obtain reliable symbols for the control and coding analysis.

Write the unit-magnitude reference as $R_{s,k}=\exp(j\pi\rho_{s,k}/2)$, where $\rho_{s,k}\in\{0,1,2,3\}$ specifies the rotation at slot $s$ and carrier $k$. The saved reference uses binary components with recurrence
\begin{equation}\label{eq:pn15component}
p_{\mathrm{PN}}[n]=p_{\mathrm{PN}}[n-14]\oplus p_{\mathrm{PN}}[n-15].
\end{equation}
Its nonzero-seed period is 32767. To state the alignment unambiguously, we initialize $p_{\mathrm{PN}}[0],\ldots,p_{\mathrm{PN}}[14]$ with the bit string \texttt{000000100000010}. Define $a(k)$ as the zero-based ordinal among the 1020 loaded carriers, from $a(-512)=0$ to $a(511)=1019$. This ordinal includes pilots and skips the four central nulls. 

The two binary components that select a four-phase rotation are aligned to this PN15 sequence separately. We compare the known pilot components with cyclic shifts of the generated sequence, accounting for each pilot's carrier ordinal. The saved alignment starts this comparison at slot 17. At that slot and ordinal zero, the two recovered sequence indices are 16327 and 16320, which specify positions within the 32767-bit PN period.

For a general resource, define $d_{s,k}=1020(s-17)+a(k)$. In the saved convention, the reference components and phase label are
\begin{equation}\label{eq:pn_grid_alignment}
\begin{aligned}
h_{s,k}&=p_{\mathrm{PN}}[(d_{s,k}+16327)\bmod32767],\\
l_{s,k}&=1\oplus p_{\mathrm{PN}}[(d_{s,k}+16320)\bmod32767],\\
\rho_{s,k}&=(2h_{s,k}+l_{s,k}+1)\bmod4.
\end{aligned}
\end{equation}
Thus moving to the next loaded carrier advances both sequence indices by one, and moving to the same carrier in the next slot advances them by 1020. At $(s,k)=(17,-512)$ the two sampled positions are 16327 and 16320. At the first negative-edge pilot, $k=-496$ and $a(k)=16$, they are 16343 and 16336. The complement in $l_{s,k}$ and the final $+1$ specify the saved bit-label and constant $90^{\circ}$ receiver-phase conventions. The phase-label parity of the SSS in slot 1 and the known edge-pilot symbols provide complementary checks of the PN-sequence alignment. The saved pilot model uses slots 17--31 for its initial component values and slot 32 to select the constant binary bias. With those choices fixed, it predicts all 4304 pilot labels over slots 33--301 without disagreement.

For the leading region, $m$ indexes a frame, $b_{m,s,k}$ is a candidate binary label, and $R_{s,k}$ is the defined unit-magnitude PN phase reference. The testable model is
\begin{equation}\label{eq:pnmodel}
\widehat X_{m,s,k}\simeq R_{s,k}(1-2b_{m,s,k}),\quad b_{m,s,k}\in\{0,1\}.
\end{equation}
Here the leading region means the contiguous non-pilot interval that starts at $(2,0,-512)$ and follows the binary-symbol model in~\eqref{eq:pnmodel}. We stop at its transition to a region that no longer follows this model, rather than combining later, separate BPSK intervals. Table~\ref{tab:leadingranges} lists this interval for each of the 18 frames. Its exact endpoints are checked together with the later coding and allocation analysis.

\begin{table}[!t]\centering\footnotesize
\caption{Contiguous leading BPSK$\times$PN intervals in the 18 frames. Endpoints are inclusive. Pilots and DC nulls are excluded.}\label{tab:leadingranges}
\setlength{\tabcolsep}{3pt}
\begin{tabular}{@{}lccr@{}}\toprule
Frame & Start $(s,q,k)$ & End $(s,q,k)$ & Symbols\\\midrule
R00 & $(2,0,-512)$ & $(14,695,+195)$ & 12744 \\
R01 & $(2,0,-512)$ & $(16,735,+235)$ & 14792 \\
R02 & $(2,0,-512)$ & $(12,883,+383)$ & 10924 \\
R03 & $(2,0,-512)$ & $(10,615,+115)$ & 8648 \\
R04 & $(2,0,-512)$ & $(14,695,+195)$ & 12744 \\
R05 & $(2,0,-512)$ & $(28,975,+475)$ & 27080 \\
R06 & $(2,0,-512)$ & $(12,769,+269)$ & 10810 \\
R07 & $(2,0,-512)$ & $(14,695,+195)$ & 12744 \\
R08 & $(2,0,-512)$ & $(43,251,-253)$ & 41416 \\
R09 & $(2,0,-512)$ & $(12,769,+269)$ & 10810 \\
R10 & $(2,0,-512)$ & $(10,615,+115)$ & 8648 \\
R11 & $(2,0,-512)$ & $(14,695,+195)$ & 12744 \\
R12 & $(2,0,-512)$ & $(28,975,+475)$ & 27080 \\
R13 & $(2,0,-512)$ & $(29,199,-305)$ & 27308 \\
R14 & $(2,0,-512)$ & $(12,769,+269)$ & 10810 \\
R15 & $(2,0,-512)$ & $(26,935,+435)$ & 25032 \\
R16 & $(2,0,-512)$ & $(8,575,+75)$ & 6600 \\
R17 & $(2,0,-512)$ & $(8,575,+75)$ & 6600 \\
\bottomrule\end{tabular}
\end{table}

\begin{figure}[t]\centering
\subfloat[]{\includegraphics[width=.48\columnwidth]{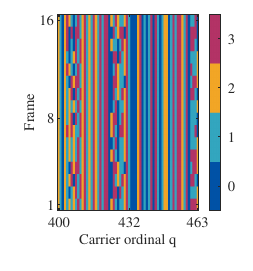}}\hfill
\subfloat[]{\includegraphics[width=.48\columnwidth]{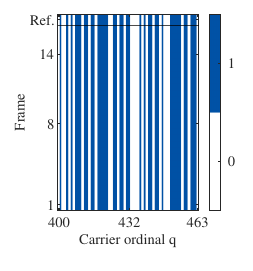}}
\caption{Phase labels and reference-parity agreement in slot 2, $q=400$--463, across 16 illustrative frames.}\label{fig:phasegrid}
\end{figure}

Changing the binary value reverses the symbol by $\pi$ while preserving whether it lies on the real or imaginary axis. With measured phase label $\ell_{m,s,k}$, this predicts
\begin{equation}\label{eq:phaseparity}
\ell_{m,s,k}\bmod2=\rho_{s,k}\bmod2.
\end{equation}
Fig.~\ref{fig:phasegrid} compares measured labels and their parity at identical resource positions across frames. The reference is fixed by its recurrence and alignment, rather than fitted separately to every observed symbol.

All 287534 tested leading-region symbols across the 18 frames satisfy the predicted parity. After reference removal, their BPSK EVM is 0.1329--0.1536. In R02 the leading binary region spans $(2,0,-512)$ to $(12,883,+383)$, comprising 10924 non-pilot symbols. Its length varies from 6600 to 41416 across the capture, as listed in Table~\ref{tab:leadingranges}. 

Unless explicitly stated otherwise, the subsequent modulation and coding analysis uses the reference-removed observations
\begin{equation}\label{eq:reference_removed}
\widetilde X_{s,k}=\widehat X_{s,k}R_{s,k}^{*}.
\end{equation}
The reference is applied to all analyzed non-pilot resources in slots 2--301, not just the binary region. After equalization and PN phase derotation, the leading-region symbols cluster around two opposite points on the real axis. Their real-part signs provide hard decisions on the candidate coded bits used for the convolutional-code and CRC analysis in Section III. To examine possible block sizes, we test 2048 symbols as a candidate long unit. Dividing the leading-region lengths by this unit leaves remainders 456, 570, and 684, spaced by 114 symbols. This motivates testing a short 114-symbol interval. The length pattern alone establishes neither size as a coding block. Sections III and V test the short and long candidates through their coding relations.

\section{Terminated Control Groups and Coding Schemes}
\subsection{Convolutional Generator Polynomials}

\begin{figure}[t]\centering
\includegraphics[width=\columnwidth]{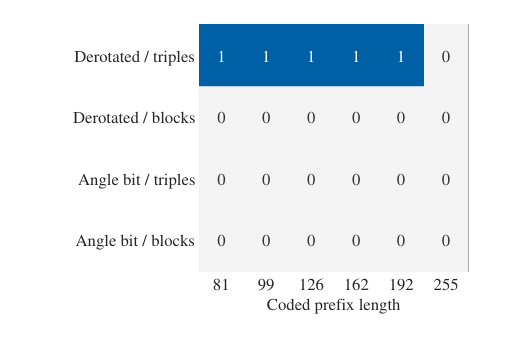}
\caption{Generator-equation nullity for the tested bit labels and output orderings at offset zero.}\label{fig:convdiscovery}
\end{figure}

The leading BPSK region identified in Section II-C provides candidate coded bits after equalization, PN phase derotation, and hard decisions on the real components. We examine these bits from $(2,0,-512)$ before attempting information-bit recovery. The rate-1/3, non-recursive convolutional header encoder described in US12074683B1 has six memory bits and 64 states~\cite{mccormick2024ofdm}. This motivates a candidate with three outputs per input and seven generator taps. The generator coefficients, serialization order, and coding-block length remain to be determined from the received bits.

For each fitting frame $f$, let $d_f[i]$ be the candidate bit at position $i$ within a selected window of $3J$ bits. Thus $J$ is the number of bits in each of three candidate output streams. With output index $a\in\{0,1,2\}$ and stream-time index $n=0,\ldots,J-1$, interleaved serialization gives
\[
c_{a,f}[n]=d_f[3n+a].
\]
The alternative $c_{a,f}[n]=d_f[aJ+n]$ assigns three consecutive equal-length segments to the outputs. These hypotheses correspond to serializing the three outputs at each input time or transmitting each output stream as a complete segment. Both are tested using reference-corrected sign bits and, separately, binary labels extracted from the four-phase angles. Window starts are shifted by 0--16 coded bits to assess alignment sensitivity. The tested window lengths are 81, 99, 126, 162, 192, and 255 bits. These finite search settings do not specify coding-block boundaries.

To identify the taps without knowing the input, let $\xi$ denote a unit delay, $u_f(\xi)$ the input polynomial in frame $f$, and $c_{a,f}(\xi)$ the polynomial representation of stream $c_{a,f}[n]$. The candidate generator is $g_a(\xi)=\sum_{\ell=0}^{6}g_{a,\ell}\xi^\ell$, where $g_{a,\ell}\in\{0,1\}$ is the coefficient at delay $\ell$. For ideal encoder outputs, $c_{a,f}(\xi)=u_f(\xi)g_a(\xi)$. Cross-convolving two outputs eliminates $u_f$ and yields
\begin{equation}\label{eq:convdual}
g_b(\xi)c_{a,f}(\xi)\oplus g_a(\xi)c_{b,f}(\xi)=0.
\end{equation}
Here multiplication denotes binary convolution, and $\oplus$ denotes XOR. Received candidate bits may differ from the encoder's binary polarity convention. The fitted relation therefore includes a pairwise binary constant $\beta_{ab}$:
\[
\bigoplus_{\ell=0}^{6}\left[g_{b,\ell}c_{a,f}[n-\ell]
\oplus g_{a,\ell}c_{b,f}[n-\ell]\right]\oplus\beta_{ab}=0.
\]
This equation is imposed for $(a,b)=(0,1),(0,2),(1,2)$ and $n=6,\ldots,J-1$, where all seven received positions are available. For each output pair, a single bias \(\beta_{ab}\) is fitted jointly across all tested time indices and frames. The measured candidate bits form the equation coefficients, and the unknowns are the 21 generator taps and three pairwise biases. The coefficient matrix can be define as the seven-bit row
\[
\mathbf r_{a,f}[n]=[c_{a,f}[n],\ldots,c_{a,f}[n-6]]
\]
and the coefficient column $\mathbf g_a=[g_{a,0},\ldots,g_{a,6}]^{\mathsf T}$, where $\mathsf T$ denotes transpose. Order the unknowns as
\[
\boldsymbol\theta=
[\mathbf g_0^{\mathsf T},\mathbf g_1^{\mathsf T},
\mathbf g_2^{\mathsf T},\beta_{01},\beta_{02},\beta_{12}]^{\mathsf T}.
\]
For one frame and time index, the three output-pair equations have coefficient block
\[
B_f[n]=\left[\begin{array}{cccccc}
\mathbf r_{1,f}[n]&\mathbf r_{0,f}[n]&\mathbf0_7&1&0&0\\
\mathbf r_{2,f}[n]&\mathbf0_7&\mathbf r_{0,f}[n]&0&1&0\\
\mathbf0_7&\mathbf r_{2,f}[n]&\mathbf r_{1,f}[n]&0&0&1
\end{array}\right],
\]
where $\mathbf0_7$ is a seven-element zero row. The three rows correspond to pairs $(0,1)$, $(0,2)$, and $(1,2)$, in that order. Each row contains three seven-column generator blocks followed by three bias columns, so $B_f[n]\in\mathrm{GF}(2)^{3\times24}$. The joint matrix is the vertical stack
\[
A_{\rm conv}=\left[\begin{array}{c}
B_{\mathrm{R00}}[6]\\ \vdots\\ B_{\mathrm{R00}}[J-1]\\
B_{\mathrm{R02}}[6]\\ \vdots\\ B_{\mathrm{R02}}[J-1]\\
B_{\mathrm{R03}}[6]\\ \vdots\\ B_{\mathrm{R03}}[J-1]
\end{array}\right].
\]
Consequently, the joint estimation problem is
\[
A_{\rm conv}\boldsymbol\theta=\mathbf0,
\qquad A_{\rm conv}\in\mathrm{GF}(2)^{9(J-6)\times24}.
\]
For an 81-bit window, $J=27$ gives $A_{\rm conv}\in\mathrm{GF}(2)^{189\times24}$. Binary Gaussian elimination determines the nullspace of this received-bit matrix~\cite{marazin2011}. The zero vector is excluded because it defines no generator.

Fig.~\ref{fig:convdiscovery} compares the resulting nullspace dimensions at zero offset. The reference-corrected bits in interleaved order give rank 23 for each tested length 81, 99, 126, 162, and 192. All five matrices have the same one-dimensional nullspace and hence the same unique nonzero solution over GF(2). The other displayed label and ordering combinations have no nonzero solution. The first 21 components of the accepted solution yield
\[
\begin{array}{ccl}
a & (g_{a,6}\cdots g_{a,0}) & g_a(\xi)\\
0 & 1101101 & 1+\xi^2+\xi^3+\xi^5+\xi^6\\
1 & 1010011 & 1+\xi+\xi^4+\xi^6\\
2 & 1011111 & 1+\xi+\xi^2+\xi^3+\xi^4+\xi^6
\end{array}
\]
The tap strings are written from delay six to zero and correspond to the octal generators $(155,123,137)_8$. The remaining components give $(\beta_{01},\beta_{02},\beta_{12})=(1,1,0)$ for the candidate-bit convention. 

Extending the window to 255 coded bits eliminates the common nonzero solution. However, shifting the window start by multiples of three can preserve the recovered taps because the three-output ordering remains unchanged. The generator relations therefore do not uniquely determine the coding-block boundaries.

\subsection{Coding Blocks and Group Boundaries}

\begin{table*}[t]\centering\footnotesize
\caption{Distribution of the two candidate block groups across 18 frames. Endpoints are inclusive and exclude pilots and DC nulls.}\label{tab:candidategroupdistribution}
\setlength{\tabcolsep}{3pt}
\begin{tabular}{@{}lccrrrp{.31\textwidth}@{}}\toprule
Group & Start $(s,q,k)$ & End $(s,q,k)$ & Blocks & Resources & Frames & Frame identifiers\\\midrule
First & $(2,0,-512)$ & $(2,227,-277)$ & 2 & 228 & 18 & R00--R17\\
Second & $(8,348,-156)$ & $(8,575,+75)$ & 2 & 228 & 4 & R03, R10, R16, R17\\
Second & $(12,428,-76)$ & $(12,769,+269)$ & 3 & 342 & 3 & R06, R09, R14\\
Second & $(12,428,-76)$ & $(12,883,+383)$ & 4 & 456 & 2 & R02, R13\\
Second & $(14,468,-36)$ & $(14,695,+195)$ & 2 & 228 & 9 & R00, R01, R04, R05, R07, R08, R11, R12, R15\\
\bottomrule\end{tabular}
\end{table*}

With the generator taps and output convention fixed in Section III-A, we now test whether the received bits form independently terminated coding blocks. Viterbi decoding~\cite{viterbi1967,forney1973} supplies candidate input bits and states. We then encode those input bits again and compare every output with the observed candidate bits. The check requires both a valid trellis path and an exact output match, rather than accepting a plausible-looking decoded word.

The 114-symbol spacing discussed in Section II-C provides a candidate length to test. After reference-phase removal, each BPSK symbol gives one candidate bit. Starting at $(2,0,-512)$, we therefore examine a 114-bit window and the immediately following 114-bit window. With the recovered rate-1/3 encoder, each window corresponds to $114/3=38$ input bits. We test whether each window follows the encoder exactly from a zero initial state, ends in a zero state, and has six zero inputs at its end. Six zeros are a natural termination candidate because they clear the recovered encoder's six-bit memory. Both windows pass these checks in R00, R02, and R03 and contain nonzero retained content. Removing the verified tails leaves $38-6=32$ retained bits per block. Thus the 32-bit word follows from testing the proposed 114-bit output interval, rather than being assumed beforehand. 

We call each independently terminated 114-bit interval that passes the state, tail, and re-encoding checks a coding block. In R02, two adjacent coding blocks occupy slot 2, $q=0$--113 and $q=114$--227. Together they span $(2,0,-512)$ through $(2,227,-277)$ and contain 228 coded resources. We provisionally call this consecutive set the first candidate block group. Applying the same encoder, state, tail, and re-encoding checks to later positions reveals four consecutive coding blocks in slot 12, from $(12,428,-76)$ through $(12,883,+383)$. These 456 resources form R02's second candidate block group, shown in Fig.~\ref{fig:controlinputs}(d). The scan proposes starting positions using observed first-word patterns and advances in 114-symbol steps until a block fails its checks or decodes to an all-zero retained word. Table~\ref{tab:candidategroupdistribution} summarizes the two candidate block groups in all 18 frames. Every first group has two blocks at the same slot-2 coordinates. The second group occurs in slot 8, 12, or 14 and contains two, three, or four blocks.

With the generators, bit polarity, output order, and state convention frozen from R00, R02, and R03, the first groups in the other 15 frames reproduce all $15\times228=3420$ coded symbols exactly and satisfy the zero-state and six-zero-tail checks. 

\subsection{CRC Polynomial}

\begin{figure}[t]\centering
\subfloat[]{\includegraphics[width=1\columnwidth]{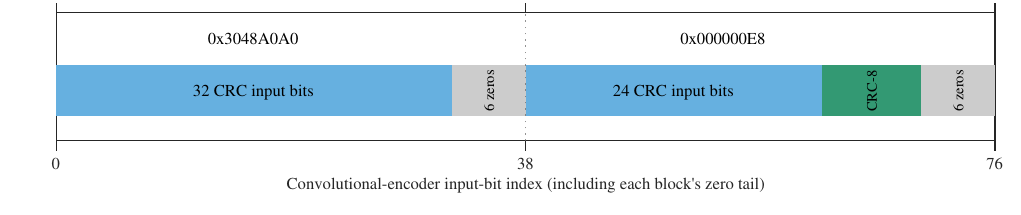}}\\[2pt]
\subfloat[]{\includegraphics[width=1\columnwidth]{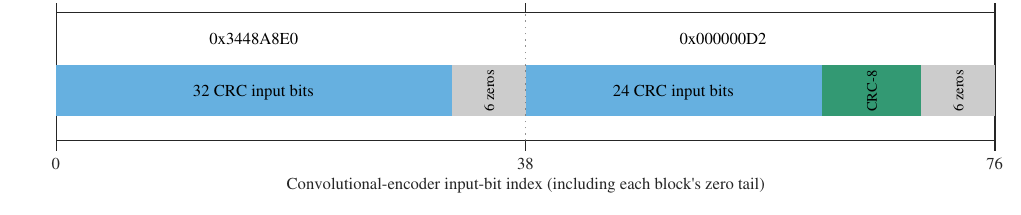}}\\[2pt]
\subfloat[]{\includegraphics[width=1\columnwidth]{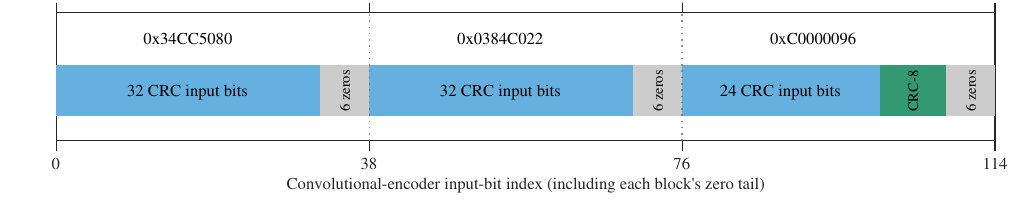}}\\[2pt]
\subfloat[]{\includegraphics[width=1\columnwidth]{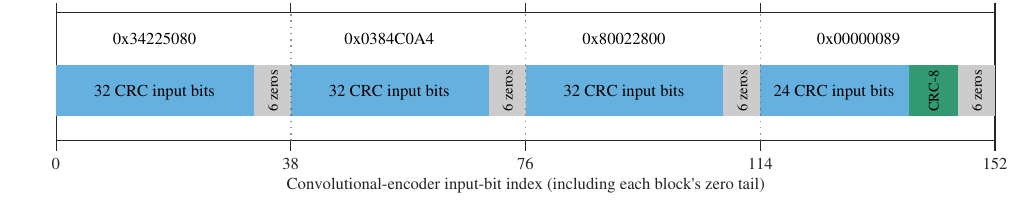}}
\caption{Convolutional-encoder inputs: (a) R02 first group, 228 resources, (b) R00 second group, 228 resources, (c) R06 second group, 342 resources, and (d) R02 second group, 456 resources. Blue denotes CRC input, green the CRC byte, and gray each block's zero tail.}\label{fig:controlinputs}
\end{figure}

The preceding analysis identifies candidate groups of two, three, or four independently terminated coding blocks. After removing each block's six zero tails, we concatenate its 32 retained bits in resource order. The remaining question is whether adjacent blocks form one information unit. The last word of each candidate group has a changing low byte, which motivates testing those final eight bits as a check value predicted from the preceding bits. We compare two input definitions: one preceding 32-bit word, and the retained bits of the complete group. For the single-word definition, we compare degree-eight polynomials, bit and byte orders, and input slices, requiring exact check-byte agreement across the tested word/byte pairs. No candidate satisfies all of those comparisons. We therefore test whether the check byte depends on the complete group rather than on one word alone.

Fig.~\ref{fig:controlinputs} shows the input range supported by the subsequent tests. The blue segments are concatenated for the CRC calculation, the green final byte is predicted, and the gray zero tails are excluded. For $N_b$ coding blocks, the input contains $32N_b-8$ bits, giving 56, 88, and 120 bits for two-, three-, and four-block groups. Panels (a) and (d) show R02's two groups, while panels (b) and (c) illustrate the shorter second groups in R00 and R06. The final block's 32 retained bits contain 24 CRC input bits and the eight-bit check value. Together with its six zero tails, these 38 inputs produce 114 coded outputs. The colors classify encoder inputs, not independently transmitted uncoded fields. Any information-layer padding remains part of the CRC input.

The search compares byte order, bit order, reflection conventions, and whether the final word's upper 24 bits enter the calculation. For the full-input, most-significant-bit-first convention, it enumerates 255 nonzero eight-bit polynomial masks and 256 initial register states, requiring one common final XOR to explain groups of different lengths. The six groups from R00, R02, and R06 leave one parameter tuple in this search: mask \texttt{0x4D}, initial register \texttt{0xFF}, final XOR zero, and no input or output reflection. Including the implicit highest-order term, the polynomial is
\begin{equation}\label{eq:controlcrc}
g_{\mathrm{CRC}}(x)=x^8+x^6+x^3+x^2+1.
\end{equation}
For each input bit, its XOR with the register's most-significant bit determines the feedback. The register shifts left within eight bits and is XORed with \texttt{0x4D} when the feedback is one. Its final value is compared with the received terminal byte. This specifies the tested CRC directly, without relying on a catalogue name. CRC performance depends on message length~\cite{koopman2004}, but no optimality claim is required here.

For R02's first group, the decoded words are \texttt{0x3048A0A0} and \texttt{0x000000E8}. The first word and the final word's upper 24 bits give 56 CRC input bits and predict \texttt{0xE8}, matching the final byte. For R02's four-block second group, the words \texttt{0x34225080}, \texttt{0x0384C0A4}, \texttt{0x80022800}, and \texttt{0x00000089} give 120 input bits and predict \texttt{0x89}. The check byte itself is convolutionally encoded. In the final block, its eight input bits generate 24 outputs, followed by 18 outputs during the six zero-tail inputs.

\begin{figure}[t]\centering
\includegraphics[width=\columnwidth]{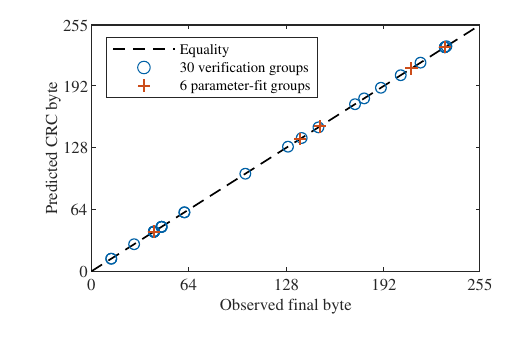}
\caption{Predicted and observed CRC bytes for fitting and verification groups.}\label{fig:crcdiscovery}
\end{figure}

We next freeze the CRC input rule and parameters selected from the six fitting groups. All 30 remaining groups in the other 15 frames produce the observed check byte, including 14 distinct group contents absent from the fitting set. Fig.~\ref{fig:crcdiscovery} compares predicted and observed bytes for both sets. Markers can overlap when bytes repeat. Across all 18 frames, the 36 groups have zero CRC differences. Thus the polynomial explains changing contents and three group lengths beyond the examples used to select its parameters, rather than merely matching one terminal byte.

The CRC results support treating each candidate block group as one checked unit. Each of the 18 frames contains two such units in resource order. Following the protocol data unit (PDU) header description in US12074683B1, we refer to these units as the first PDU header and the second PDU header. The first always occupies 228 resources in slot 2, while the second changes position and contains two, three, or four coding blocks. Section IV shows how fields in these headers correspond to the modulation and lengths of following allocations, which supports the control interpretation.

\section{Control-Field and Resource Allocation}
\label{sec:allocation}
\subsection{PDU Header Prefix and Field Layout}
\label{sec:layouthypotheses}
\begin{figure*}[t]\centering
\includegraphics[width=.98\textwidth]{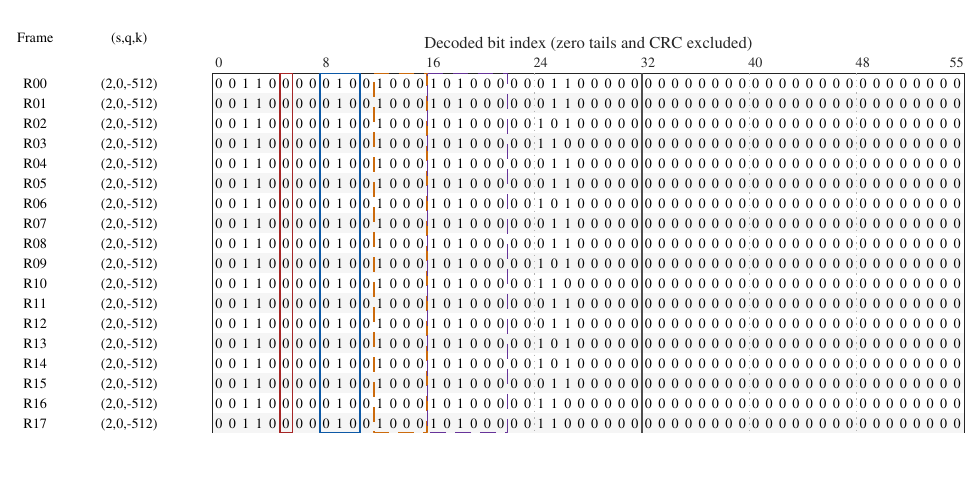}
\caption{CRC-input bits of the first PDU header in all 18 records. Solid red/blue boxes mark the observed role/block-count associations. Dashed orange/purple boxes mark the candidate entry-count/format-code windows.}\label{fig:controlbitsfirst}
\end{figure*}

\begin{figure*}[!t]\centering
\subfloat[]{\includegraphics[width=.98\textwidth]{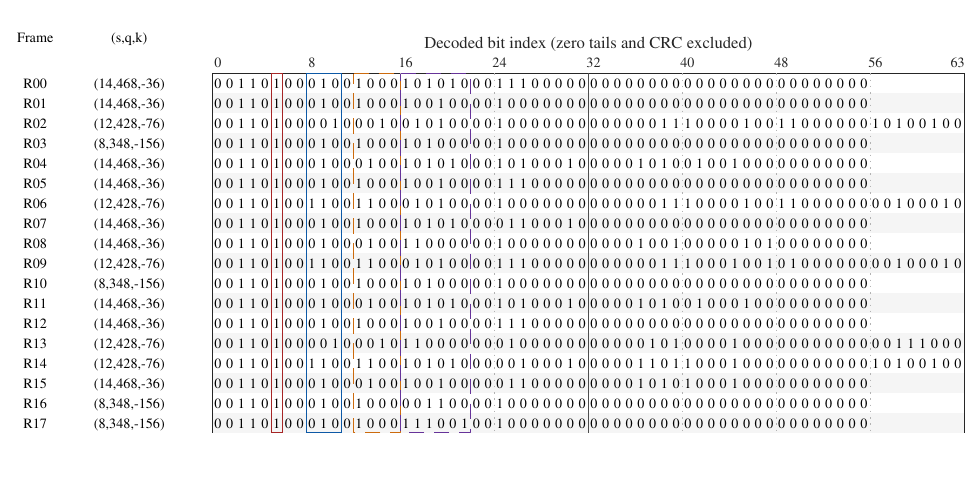}}\\[2pt]
\subfloat[]{\includegraphics[width=.98\textwidth]{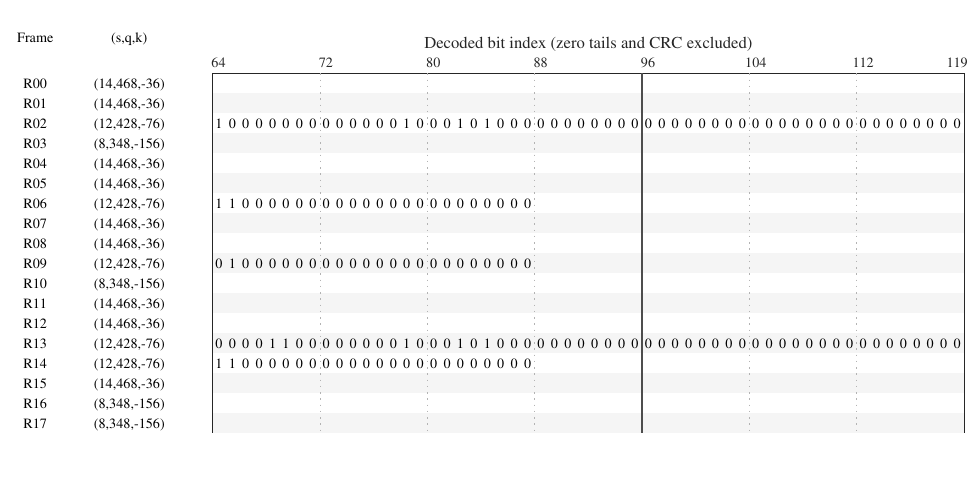}}
\caption{CRC-input bits of the second PDU header: (a) positions 0--63 and (b) positions 64--119. Blank cells lie outside a row's CRC-input interval. Box styles follow Fig.~\ref{fig:controlbitsfirst}.}\label{fig:controlbitssecond}
\end{figure*}

Section III establishes two CRC-checked PDU headers in each frame. A header containing $N$ independently terminated coding blocks retains $32N$ bits after the six zero tail bits of each block are removed. The final eight retained bits form the CRC. Figures~\ref{fig:controlbitsfirst} and~\ref{fig:controlbitssecond} compare the other $32N-8$ bits in all 18 frames. The displayed positions start at zero in the first retained word. Blank cells lie beyond a header's CRC-input interval and do not represent zero bits.

The first PDU headers have exactly the same first 16 bits, \texttt{0011000001001000}, in all 18 frames. Their lengths are also identical: two coding blocks, or 228 coded resources. The second PDU headers instead have four patterns in these positions, occurring in 9, 4, 3, and 2 frames. Their independently decoded lengths are 228, 342, or 456 coded resources. This contrast suggests that the beginning contains information about the header itself, before the bits that describe the following resources.

Two associations can be checked without interpreting the following payload. Bit 5 is zero in every first header and one in every second header. Bits 8--10 read \texttt{010}, \texttt{110}, or \texttt{001}. Assigning increasing weights $1,2,4$ from left to right gives $N=2,3,4$, matching the independently recovered block counts. For example, R06's second header has three blocks and R02's has four. These observations support a header-role indicator and a block-count field. The bit weights are a declared receiver numbering convention. The observations do not establish standardized field names.

Under the declared bit-weight convention, bits 12--15 give values of 1 in every first PDU header and 1--4 in the second PDU headers. We test whether this value specifies the number $D$ of following entries. This count is not the block count $N$: R08's second header has $D=2$ and $N=2$, whereas R02's first header has $D=1$ and $N=2$. Across all 36 headers, the pairs $(D,N)=(1,2),(2,2),(3,3),(4,4)$ occur 27, 4, 3, and 2 times. We therefore test a layout with a 16-bit prefix, $D$ equal-width entries of $W$ bits, an unassigned interval of $P_{\mathrm{pad}}$ bits, and the known CRC:

\begin{equation}\label{eq:entrypacking}
16+DW+P_{\mathrm{pad}}+8=32N.
\end{equation}
This is a layout hypothesis, not a conclusion from constant bits alone. Prefix constancy and the associations with header length motivate the separation, but cannot uniquely place its boundary at bit 16 or prove that $D$ counts allocations. The next subsection tests the proposed boundary and entries against both the retained bits and the received signal.

\subsection{Format of PUD Header}
\label{sec:resourcecounts}
We first test the entry width within the layout proposed in Section IV-A. For each $W=8,\ldots,32$, the residual interval must contain a nonnegative number of bits, all its observed bits must be zero, and $\lceil(16+DW+8)/32\rceil=N$ must hold. The last condition assumes the minimum number of retained 32-bit words. Only $W=19$ and $W=20$ satisfy these tests across all 36 PDU headers. The CRC covers the concatenated retained bits and therefore does not distinguish the two candidate entry boundaries. Residual zeros may be padding or reserved fields.

The first six entry bits provide an initial association with modulation. The first entry begins at retained-bit position 16 in either candidate layout. In R02, these bits are \texttt{101000} in the first header and \texttt{010100} in the second. Local comparisons over the next 256 equalized symbols favor BPSK and QPSK, respectively, corresponding to regions 2 and 4 of Fig.~\ref{fig:atlasR02}. R17's second header begins with \texttt{111001}, followed by the sixteen-point observations in region 4 of Fig.~\ref{fig:atlasR17}. With increasing weights $2^0,\ldots,2^5$ from left to right, these patterns give $2^0+2^2=5$, $2^1+2^3=10$, and $2^0+2^1+2^2+2^5=39$. These regional observations associate the low six bits with format, but do not establish the full index width or identify a modulation boundary from an individual symbol. 

The BPSK interval between the two PDU headers provides a simpler test of the quantity bits because its modulation is unchanged across the frames and its limits follow from the PDU header positions recovered in Section III. In a 20-bit working entry, the last twelve positions are retained-bit positions 24--35 of the first header. Their values in R00, R02, and R03 are \texttt{011000000000}, \texttt{101000000000}, and \texttt{110000000000}. Reading these with weights $2^0,\ldots,2^{11}$ gives 6, 5, and 3. The corresponding independently counted BPSK regions contain 12288, 10240, and 6144 non-pilot symbols, shown as region 2 of those frames in Appendix~\ref{app:regions}. Each measured length is 2048 times the value read from the twelve positions. This motivates a candidate resource-unit count $C$ at these positions. The comparison extends to every first PDU header: all 18 count values predict the intervening code-5 region exactly. These small values do not establish the full count-field width or, by themselves, show that the units are codewords.

The patent description provides a functional interpretation to test. US12074683B1 states that the MCS field specifies encoding and modulation, while the MCS length field specifies payload size~\cite{mccormick2024ofdm}. This supports looking for a format index and a quantity within each candidate MCS index--length pair, without fixing their widths or identifying the quantity's unit in this capture. We use MCS index for the candidate format identifier and MCS length for its associated quantity field. In the tested codeword model, MCS length specifies a codeword count, not a byte count. This terminology follows the functional distinction in the patent without asserting identical field definitions. Separately, the example MCS table in Fig. 4 of US11792832B1 includes indices 125 and 126~\cite{iyer2023amc}. Direct binary representation of 126 requires at least seven bits. The figure is an illustrative subset, so 126 is not an established maximum index. Nor does that example prove that the present control words use the same complete index space.

Combining these clues, we retain two candidate interpretations of a 20-bit MCS index--length pair. Number its received bits from 0 to 19. The first uses bits 0--6 as a seven-bit MCS index, bit 7 as an unresolved bit, and bits 8--19 as the candidate MCS length $C$. The second uses bits 0--7 as an eight-bit MCS index and bits 8--19 as the same MCS length. MCS index and MCS length fields are each read with increasing weights starting at $2^0$. For the first MCS index--length pair of a header, these positions correspond to retained-bit positions 16--22, 23, and 24--35 in the first interpretation, or 16--23 and 24--35 in the second. Both layouts place the quantity at the positions associated with the measured BPSK lengths.

All 52 MCS index--length pairs in the 36 PDU headers have a zero eighth bit (position 7). Consequently, the seven-bit and eight-bit candidates produce identical indices and quantities throughout this capture. The observations cannot distinguish a seven-bit index followed by an unresolved zero bit from an eight-bit index with a zero most significant bit. Under either reading, the MCS index--length pair previously named by its low-six-bit label 4 is instead indexed as 68. Its received pattern is
\[
\begin{array}{ll}
\texttt{0010001}\;|\;\texttt{0}\;|\;\texttt{100000000000}
 & (7+1+12),\\
\texttt{00100010}\;|\;\texttt{100000000000}
 & (8+12).
\end{array}
\]
The index is $2^2+2^6=68$ and the displayed quantity is one. This relabeling affects four MCS index--length pairs in R02, R06, R09, and R13, with quantities 1, 3, 2, and 1. We use 68 hereafter for that allocation family. The other observed format numbers are unchanged. Preliminary six-bit labels in the preceding sections and labels retained in evidence filenames are related by this explicit mapping, not by a change in the received resources. Index 68 has no corresponding row in the published example subset, and its encoder remains unidentified.

The second PDU headers test whether the same quantity interpretation extends to multiple MCS index--length pairs and different modulations. R02 gives indices and quantities $(10,1)$, $(28,50)$, $(37,1)$, and $(68,1)$ for its subsequent regions 4--7. Both candidate layouts give these same values. In R08, the different indices 3 and 9 describe two BPSK regions, showing that a modulation change is not required between consecutive MCS index--length pairs. Thus the MCS index--length pairs are interpreted through their numerical and coding relations as well as their constellations. The complete 18-frame bit-by-bit comparison supplies 18 first-header and 34 second-header MCS index--length pairs, not independent acquisitions.

To predict an allocation's symbol length from $C$, its format must also supply a candidate codeword length and modulation mapping. Let $N_c$ be the coded bits in a candidate codeword, $b_c$ the coded bits mapped to each symbol, $Q_c$ the symbols per codeword, and $S_c$ the symbols in the allocation. If each codeword begins on a new symbol, the prediction is $S_c=CQ_c$. Concatenating all coded bits before mapping instead gives the competing length $S_{\mathrm{joint}}$:
\begin{equation}\label{eq:round}
\begin{aligned}
Q_c&=\left\lceil\frac{N_c}{b_c}\right\rceil,\qquad S_c=CQ_c,\\
S_{\mathrm{joint}}&=\left\lceil\frac{CN_c}{b_c}\right\rceil.
\end{aligned}
\end{equation}
The ceiling accounts for indivisible modulation symbols. Under separate allocation, spare capacity remains in each codeword's last symbol. For BPSK code 5, the candidate $N_c=2048$ and $b_c=1$ yield the 2048-symbol unit found above. Section V tests the encoder relation supporting its codeword interpretation. For other formats, neither $N_c$ nor $b_c$ is determined by the count bits alone.

R02's second-header code-28 quantity $C=50$ distinguishes the two allocation rules. With candidate $N_c=16384$ and $b_c=3$, each word requires $Q_c=5462$ symbols because 5461 symbols hold only 16383 bits. Separate allocation predicts $50\times5462=273100$ symbols, whereas joint mapping predicts 273067, a difference of 33. The former places the candidate region 5 from $(21,40,-464)$ through $(293,51,-453)$. These positions are predictions to validate, not an independently established modulation boundary or a complete code-28 encoder.

\begin{figure}[t]
\centering\includegraphics[width=\columnwidth]{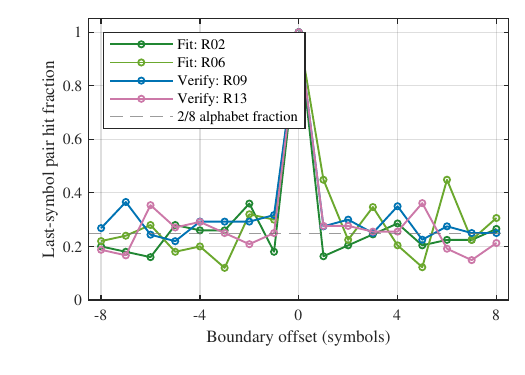}
\caption{Frozen-pair hit fractions at candidate code-28 last symbols and shifted positions. The dashed $2/8$ line is an alphabet fraction.}\label{fig:endpoints}
\end{figure}

We first test the proposed internal spacing without using the overall allocation endpoint as the only criterion. At intervals of 5462 symbols from the candidate field start, we collect the symbol predicted to end each codeword. A reproducible restriction on these symbols would support the proposed spacing. The 100 candidate words in R02 and R06, 50 in each frame, select a pair of points from the observed eight-point alphabet. We freeze that pair and test the 41 candidate words in R09 and 48 in R13. All 89 predicted last symbols lie in the selected pair. Figure~\ref{fig:endpoints} identifies the fitting and verification curves by these frame numbers.

To check whether the point restriction merely reflects a general preference for those points, we move each tested last-symbol position by $-8,\ldots,8$ symbols. The horizontal axis of Fig.~\ref{fig:endpoints} is this displacement, and the vertical axis is the fraction of tested positions falling in the frozen pair. The four curves separate the two fitting frames from the two verification frames. Only zero displacement gives complete agreement in every frame. Where a shifted position leaves the saved field, it is excluded from that offset's denominator. The dashed $2/8$ line denotes two points out of eight, not a calibrated probability or significance threshold.

The repeated last-symbol restriction supports the 5462-symbol candidate spacing and an allocation rule in which each codeword uses a separate final modulation symbol. Under the 16384-bit model, that symbol has capacity for two additional bits. The analogous 8192-bit candidates for codes 37 and 27 occupy 2731 three-bit symbols, leaving one bit of capacity. These statements do not recover code 28's parity checks or prove its bit length uniquely. The point-pair test does not establish transmitter bit labels or polarity, so the actual spare-bit values remain unresolved. Section~\ref{sec:endpointvalidation} next tests the complete allocation-length prediction against a transition estimated from the received symbols.

\subsection{Allocation-Boundary Validation}
\label{sec:endpointvalidation}
The preceding subsection predicts allocation lengths from the candidate MCS indices and quantities. We now test whether their sum reaches the payload endpoint observed in the received signal. After PN phase removal, the terminal observations of each frame remain close to a constant over an extended interval. We call this behavior reference-symbol padding. The signal observation motivating this test is illustrated by region 8 of R02 in Fig.~\ref{fig:atlasR02} of Appendix~\ref{app:regions}. Since a payload symbol may also coincide with the reference, a single matching symbol cannot identify this transition. We search for the position that best explains the preceding observations as varying payload symbols and the following observations as reference symbols, using the combined score over the entire interval.

Starting immediately after the last PDU header, we order non-pilot resources by carrier within each slot and then by slot, with index $i=0$ at $(12,884,+384)$ in R02. Let $(s_i,k_i)$ identify the slot and carrier of resource $i$, and let $z_i=\widehat X_{s_i,k_i}$ be its equalized observation before PN removal. Its reference is $R_{s_i,k_i}$, using the grid defined in Section II-C. Since the reference has unit magnitude, $|z_i-R_{s_i,k_i}|=|z_iR_{s_i,k_i}^{*}-1|$. Let $t$ denote a candidate resource index for the first padding symbol. Observations with $i<t$ are explained as varying payload symbols, and those with $i\geq t$ are explained as reference symbols. We compare these explanations over the entire search interval. This avoids declaring the endpoint at the first observation close to the reference. The known PDU header boundary determines where the search starts, but no MCS-derived length restricts the candidate values of $t$.

Under a circular Gaussian model, the distance score for a candidate constellation point $a$ is proportional to $\exp(-|z_i-a|^2/\nu)$, where $\nu > 0$ is a standard squared-distance scale. By averaging over the equally weighted points in the candidate constellation $\mathcal{A}$, taking the logarithm, and selecting the highest-scoring candidate grid gives
\begin{equation}\label{eq:fillll}
\ell_V(z_i)=\max_{\mathcal A}\left[\log\sum_{a\in\mathcal A}e^{-|z_i-a|^2/\nu}-\log|\mathcal A|\right].
\end{equation}
The subtraction implements the equal-point average. The tested four-point and sixteen-point grids at candidate orientations approximate the varying observations. Selecting a grid for each observation is not a modulation decision and does not imply that the reference-removed payload uses only these two alphabets. For the reference explanation, the transmitted point is specified, giving $\ell_F(z_i)=-|z_i-R_{s_i,k_i}|^2/\nu$. For each candidate $t$, the total score is
\[
\sum_{i<t}\ell_V(z_i)+\sum_{i\geq t}\ell_F(z_i).
\]
Subtracting the all-reference score, which is independent of $t$, gives the equivalent endpoint estimate
\begin{equation}\label{eq:fillend}
\widehat t=\arg\max_t\sum_{i<t}\bigl[\ell_V(z_i)-\ell_F(z_i)\bigr].
\end{equation}

For $\nu=0.01,0.02,0.04$, the signal search places R02's first padding symbol at $(297,815,+315)$, leaving 4205 non-pilot symbols through slot 301. Only after this estimate is obtained do we compare it with the MCS prediction. Summing the predicted allocations with per-codeword rounding gives 286071 symbols after the last PDU header and reaches the same position. Across all 18 frames, 16 predicted endpoints agree exactly with the signal estimates, while the other two differ by one and two symbols.

R04 and R11 provide a further check on the second-header allocations. Under the 20-bit interpretation, the first MCS index--length pair has index 21 and quantity 69 in both frames. R00's index-21 allocation, described by one MCS index--length pair, supplies a candidate scale of 4096 symbols per count, predicting $69\times4096=282624$ symbols before the next allocation. A separate search for a QPSK-to-higher-order transition covers the final 20 slots before the reference-consistent suffix, without imposing this numerical position. The saved search gives zero measured-minus-predicted displacement at distance scales 0.01, 0.02, and 0.04. These are model-dependent transition estimates, not proof of an unambiguous modulation start.

The second MCS index--length pair is read as index 37 with quantities 1 and 2 in the 20-bit layout. Applying the same seven- or eight-bit index convention to a 19-bit MCS index--length pair instead gives index 74 with quantities 2 and 4. It does not give index 10. Consequently, a QPSK association established for index 10 cannot reject this 19-bit interpretation. The saved comparison on the next 512 symbols favors a sixteen-point grid over QPSK: at distance scale 0.02, the mean log scores are $(-12.8286,-3.5385)$ in R04 and $(-12.3721,-3.6142)$ in R11. These scores use channel-equalized symbols with pilot-based common-phase correction before per-resource PN removal. They characterize the observed region, but index 74 has no independently established modulation in this capture. The scores therefore do not uniquely exclude the 19-bit layout under the revised index convention.

The appendix plots use the reference-removed observations specified in Section II-C. In that convention, code 37 in region 5 of R04 and R11 has eight constellation points, as shown in Figs.~\ref{fig:atlasR04} and~\ref{fig:atlasR11}. Restoring the four-phase PN reference rotates these points by multiples of $\pi/2$ at different resource positions, and their pooled observations can occupy a larger sixteen-point grid. Thus the model scores and appendix plots describe different processing stages of the same signal, not competing modulation conclusions. The comparison rejects the tested QPSK explanation without establishing a complete 16QAM alphabet after reference removal, a bit mapping, or a code-37 encoder.

Together, the 36-header packing checks, quantity-to-resource correspondences, terminal-symbol tests, and 18-frame endpoint comparisons support using the conditional 20-bit working layout

\begin{equation}\label{eq:entrylayout}
\underbrace{16}_{\text{prefix}}+
\underbrace{20D}_{\text{MCS pairs}}+
\underbrace{P_{\mathrm{pad}}}_{\text{padding}}+
\underbrace{8}_{\text{CRC}}=32N.
\end{equation}
This working model retains both the seven-bit-index-plus-one-bit and the eight-bit-index interpretations. It assumes ordered contiguous allocations and consistent format meaning across MCS index--length pairs. The available comparisons do not uniquely exclude every 19-bit layout under the revised index convention. The prefix boundary, full quantity width, unresolved eighth bit, and residual-zero semantics remain model-limited. The endpoint estimate depends on the reference, recovered PDU header position, finite constellation bank, and assumed terminal transition. Overlapping candidate alphabets prevent interpreting the score as an unambiguous modulation-boundary detector. Agreement with predicted endpoints supports model consistency, not encoder identity or absence of information. Section V tests the coding relations within the allocations in Appendix~\ref{app:regions}.

\section{Channel-Code Recovery and Validation}
US12074683B1 describes LDPC encoding of payload data~\cite{mccormick2024ofdm}, providing a coding-family hypothesis for the allocations recovered in Section IV. The patent does not specify the corresponding generator or parity-check matrices. We therefore examine whether the received bits support an LDPC representation. Starting from candidate codeword boundaries and bit mappings, we recover a systematic generator and test its parity predictions on codewords excluded from fitting. We then seek a complete sparse parity-check matrix and assess consistency through decoding, re-encoding, and remodulation. 

\subsection{QC Generator Matrix Recovery}
The first task is to determine whether the candidate words share a systematic encoding relation. For a candidate codeword of length $N_c$ and dimension $K_c$, write $c=[u,p]$, where $u\in\GF^{K_c}$ and $p\in\GF^{N_c-K_c}$ are row vectors of candidate systematic and parity bits. In the declared receiver bit order, the relation is
\begin{equation}\label{eq:systematicgenerator}
p=uP,\qquad G=[I_{K_c}\mid P],\qquad c=uG.
\end{equation}
Here $P\in\GF^{K_c\times(N_c-K_c)}$ is the parity submatrix of the systematic generator matrix $G\in\GF^{K_c\times N_c}$, and $I_{K_c}$ is the identity matrix. Each column of $P$ specifies which systematic bits are XORed to generate one parity bit. All coding-matrix operations are over $\GF$, with XOR as addition.

Recovering an arbitrary $P$ requires systematic-bit observations that span all $K_c$ input dimensions. When the available codewords do not provide this span, additional structural constraints are needed. We consider a quasi-cyclic (QC) model because circulant-block structures are commonly used in LDPC codes~\cite{fossorier2004}. For a candidate lifting factor $Z$, let $T_Z$ cyclically shift every contiguous $Z$-bit block by one position in the same direction. Under the QC hypothesis, this operation produces another codeword. If the systematic-bit and parity-bit partitions align with the block boundaries, their shifted vectors must therefore satisfy the same relation $p=uP$. We assess the candidate structure through equation consistency and parity prediction on codewords excluded from generator fitting.

To construct the fitting system, we collect systematic-bit vectors from the selected codewords and their candidate QC shifts. We place these vectors in the rows of $U$ and their corresponding parity-bit vectors in the same rows of $V$. If this construction gives $M$ pairs, then $U\in\GF^{M\times K_c}$ and $V\in\GF^{M\times(N_c-K_c)}$, and the fitting problem is $UP=V$. Gaussian elimination over $\GF$ applies identical row operations to $U$ and $V$ and checks the remaining equations for consistency. A consistent system with $\operatorname{rank}U=K_c$ determines $P$ uniquely within the declared bit order and QC model. If the rank is deficient, some encoder directions remain undetermined. Inconsistent equations reject the tested combination of bit mapping, partition, and QC structure.

A fitted matrix must also explain codewords excluded from its fitting. For such a word, let $u_{\mathrm{obs}}$ and $p_{\mathrm{obs}}$ denote the observed systematic-bit and parity-bit vectors. We predict $u_{\mathrm{obs}}P$ and count disagreements using
\begin{equation}\label{eq:parerror}
e_p=\operatorname{wt}(p_{\mathrm{obs}}\oplus u_{\mathrm{obs}}P),
\end{equation}
where $\oplus$ denotes XOR and $\operatorname{wt}$ is the Hamming weight.

\subsection{Sparse Parity-Check Matrix Recovery}
A generator that predicts parity establishes a linear coding relation, but does not by itself establish an LDPC representation. The systematic generator can be dense even when the code has sparse parity checks. We therefore seek a parity-check matrix $H\in\GF^{(N_c-K_c)\times N_c}$ whose rows involve few bit positions and satisfy $Hc^\T=0$.

Searching all subsets of the $N_c$ bit positions would be prohibitive. Instead, the recovered $P$ links the two parts of each candidate check: selecting a small set of parity-bit positions determines the associated systematic-bit coefficients, or an inverse relation can be used where available. We retain candidates only when the combined number of nonzero coefficients is small. Projection hashes accelerate candidate matching, but every resulting equation is checked over all bit positions. Agreement in a projection alone is insufficient. This is a structured search for low-weight dual-code vectors~\cite{valembois2001}, rather than a solution to arbitrary noisy code recognition.

Under the accepted QC model, blockwise cyclic shifts of seed checks generate further rows of $H$. Gaussian elimination determines whether these rows provide the required $N_c-K_c$ independent checks. We verify $GH^\T=0$ and test the received codewords against the recovered checks. Together with $\operatorname{rank}G=K_c$, full check rank establishes that $G$ and $H$ describe the same code space. Isolated low-weight equations do not establish this result. 

The check matrix also provides a second way to reconstruct parity. Partition its columns as $H=[H_u\mid H_p]$, where $H_u\in\GF^{(N_c-K_c)\times K_c}$ acts on systematic bits and $H_p\in\GF^{(N_c-K_c)\times(N_c-K_c)}$ acts on parity bits. If $H_p$ is nonsingular, then
\begin{equation}\label{eq:hgenerator}
H_p p^\T=H_u u^\T
\end{equation}
uniquely reconstructs parity from the systematic bits. Comparing this prediction with $p=uP$ checks consistency between the recovered generator and parity-check representations. The sparse full-rank checks, QC structure, and parity predictions jointly support an equivalent QC-LDPC representation. They do not identify a unique transmitter implementation or establish minimum-distance, girth, or capacity claims.

\subsection{Validation by Decoding and Remodulation}
The binary matrix tests must also be consistent with the noisy received symbols and their candidate bit mappings. With the recovered encoder fixed, we compare candidate labelings of the observed constellation, decode their soft bits, and re-encode the recovered input bits. We then remodulate the predicted codewords using the inherited receiver calibration. This checks whether the coding relation and symbol mapping explain the same observations without fitting a separate target-field rotation.

For a received equalized symbol $z$, let $\mathcal A$ be the candidate constellation and $\ell_b(a)$ the bit at label position $b$ of point $a\in\mathcal A$. A Gaussian likelihood gives the log-likelihood ratio favoring bit zero over bit one:
\begin{equation}\label{eq:llr}
L_b(z)=\log\frac{\sum_{a:\ell_b(a)=0}\exp(-|z-a|^2/\sigma^2)}
{\sum_{a:\ell_b(a)=1}\exp(-|z-a|^2/\sigma^2)}.
\end{equation}
The sums run over $\mathcal A$, and $\sigma^2>0$ is an assumed likelihood scale, not an independently calibrated physical noise variance. We will test several likelihood scales to assess the sensitivity of the results to the assumption.

We employ the normalized min-sum algorithm with a normalization factor of $\eta=0.8$ as a low-complexity approximation to belief-propagation decoding~\cite{fossorier1999,chen2002}. Each check-node message uses the product of the other incoming signs and their minimum absolute value, multiplied by $\eta$. Variable-node messages combine the intrinsic likelihood with the other incoming check messages. For each tested mapping we report failed checks before decoding, changed hard decisions, parity agreement after re-encoding, and remodulation error. A zero syndrome after decoding alone is insufficient because a decoder can converge to a valid codeword under an incorrect mapping.

For a specified comparison set of symbol indices $\Omega$, let $z_i$ be the received observation and $\widehat z_i$ its remodulated prediction in the same signal coordinates. We measure
\begin{equation}\label{eq:nmse}
\mathrm{NMSE}_{\Omega}=\frac{\sum_{i\in\Omega}|z_i-\widehat z_i|^2}
{\sum_{i\in\Omega}|z_i|^2}.
\end{equation}
A parity-flip control preserves the candidate constellation while violating the recovered redundancy relation, allowing its signal error to be compared with that of the predicted codeword. The compared sets and calibration must be the same. Core-only results exclude symbols containing unexplained bits, whereas full-field results include the complete specified field. Equalized-symbol and time-domain errors use different observation sets and denominators and are not interchangeable. 

\section{Payload Coding Identification Results}
\subsection{Rate-1/2 QC-LDPC Recovery for Code 5}
R02 region 2 in Fig.~\ref{fig:atlasR02} runs from $(2,228,-276)$ to $(12,427,-77)$ and contains the five candidate code-5 words between the PDU headers. Its two-point constellation supplies binary observations. The first PDU header's count predicts the field length, while the following generator and parity tests examine its coding independently of point count.

The first BPSK body splits into 2048-bit units. Its leading 1024 bits predict the following 1024 under the accepted receiver convention. Fourteen words from R00, R02, and R03, augmented by QC128 rotations, give 1792 training equations and input rank 1024. Binary elimination yields a $1024\times1024$ parity generator with zero fitted affine offset.

Sparse dual recovery uses both parity and systematic pivot bases. For a proposed check $h=[a,b]$, the dual condition is
\begin{equation}\label{eq:sparse}
Gh^\T=a^\T+Pb^\T=0.
\end{equation}
Small parity supports propose $b$; the associated systematic support is then determined and verified in the full dimension. Complementary searches use the inverse parity generator where available. Eight seed checks and their QC128 orbits produce a rank-1024 matrix with row weight six and column weights two, three, or four. Its $8\times16$ base support appears in Fig.~\ref{fig:graphs}. The 16 block columns correspond to the 2048-bit codeword at $Z=128$. A zero entry denotes an identity circulant, whereas a missing connection denotes a zero block. Cyclically shifting a seed row can change the displayed shift values without changing the code.

The frozen predictor passes 77 reserved first-body words from 15 other frame records: 78848 parity bits disagree at zero positions. All 91 first-body words satisfy the recovered generator and checks. Two later allocations bearing the same format code also pass, giving 93 tested words in 20 fields. The representation does not identify source-byte semantics, source scrambling, or a standardized base matrix.

The three passes establish different facts: QC128-augmented equations admit a generator; the full lifted dual is sparse and complete; and the frozen representation predicts reserved observations. The raw 91 first-body words have rank at most 91, so their 1024-dimensional identifiability relies on QC, not on 1792 independent acquisitions.

The 77 words excluded from generator fitting were included in the structural screening, so they do not constitute previously unseen data for the full identification procedure. The 77 reserved words test the conditional model after fitting. Zero disagreement over 78848 parity positions is agreement against recovered bits.

\subsection{Shared Coding of Codes 10 and 39}
In R02, the first MCS index--length pair of the second PDU header points to region 4 in Fig.~\ref{fig:atlasR02}, from $(12,884,+384)$ to $(21,39,-465)$. Its 8192 QPSK symbols give 16384 candidate coded bits. The candidate codeword begins immediately after the second PDU header, rather than at an integer slot boundary. The following analysis tests its systematic relation and the ambiguity caused by insufficient input rank.

Code 10 is QPSK with $(16384,8192,1024)$. The linear fit using R02, R06, and R09 has input rank 8191, one less than the candidate information dimension of 8192. This leaves one input direction unobserved. Let a nonzero column vector $d_u\in\GF^{K_c\times1}$ define the input hyperplane $ut=0$. If $P_0$ fits the data, then
\begin{equation}\label{eq:completion}
P_v=P_0+tv^\T
\end{equation}
gives the same training predictions for any parity-side column vector $v\in\GF^{(N_c-K_c)\times1}$. The observed source relation alone does not identify a source CRC or plaintext restriction.

The saved input constraint is QC-invariant. Enforcing commutation with the QC1024 shift constrains each of the eight parity blocks in $v$ to a constant bit, leaving $2^8=256$ completions. All ten observed code-10 words lie in the same input hyperplane and therefore cannot distinguish these completions by parity agreement alone. A further declared sparse prior is essential.

The implemented sparse search uses 64 projected source coordinates, parity supports of size three or four, full check weight at most 16, and a projection-bucket cap of 64. Only completion mask 31 is compatible with the recovered rank-8192 sparse check representation. The selected matrix has row weight six and a full-rank parity submatrix. Uniqueness is within this completion family and finite sparse-search procedure; it is not arbitrary-encoder uniqueness. The R13 word is excluded from the generator fit and has zero disagreement over 8192 parity bits. Its values were, however, inspected when evaluating encoder hypotheses before the parameters were fixed. It therefore checks consistency with the fitted model rather than performance on unseen data.

The source of the 256-candidate restriction can be made explicit. The missing input direction $d_u$ is fixed by the fitting equations. Without structure, $v$ in~\eqref{eq:completion} contains 8192 free binary coordinates. Commutation with the accepted QC1024 operator requires $T_{1024}v=v$, so each of its eight contiguous blocks is constant. One binary choice per block remains. The saved input relation has support over six entire source blocks; all observed code-10 inputs satisfy it. Consequently, all 256 generators explain those words, including the R13 word excluded from fitting, before sparse checks are considered.

The sparse search selects mask 31 within its finite family; it does not remove the observed input hyperplane. A new word outside that hyperplane, if the source permits it, would directly test the missing direction. If the source always restricts inputs, observations may identify only the transmitted subspace.

Across the four code-10 fields, fixed-encoder re-encoding and remodulation give NMSE 0.01822--0.01922, compared with 1.97388--1.99058 for the parity-flip control. This comparison tests the recovered redundancy while retaining the modulation alphabet.

The single target-format field is R17 region 4 in Fig.~\ref{fig:atlasR17}.
Only R17 contains the target code-39 field. The previously recovered code-10 encoder is frozen. A 13824-candidate mapping search selects separable Gray labeling on levels $[-3,-1,1,3]/\sqrt{10}$, with single-axis map $[0,1,3,2]$ and serial order sign-I, sign-Q, amplitude-I, amplitude-Q. The best hard-decision candidate has 59 failed checks; the next has more than 2600.

At likelihood scales 0.01, 0.02, and 0.04, one decoder iteration changes 21 hard bits and returns the same word. All 8192 checks pass and all 8192 parity bits agree with the frozen encoder. The 4096-symbol field remodulation has NMSE 0.01851425; a control that flips the encoded parity bits has NMSE 1.59768657. Field gain/phase is inherited from a calibration on the first 228 PDU header symbols, not fitted on the target field. Fig.~\ref{fig:modulation} shows the received alphabet and this signal-domain contrast.

\subsection{Shared Coding of Codes 3, 12, and 27}
The second PDU header of R08 assigns code 3 to region 4 in Fig.~\ref{fig:atlasR08}, whereas the second PDU header of R16 assigns code 12 to region 4 in Fig.~\ref{fig:atlasR16}. The received constellations are BPSK and QPSK, respectively. We retain these distinct symbol-to-bit mappings when testing whether their code relations share the same graph and lifting factor.

Code 5 and code 10 expose the same column-labeled $8\times16$ incidence pattern and repeated-shift groups. Freeze those structural relations, but not numerical shifts. Within this family, one code-3 word in R08 determines the $Z=512$ shifts using its first 64 cyclic positions. Remaining 448 positions are excluded from shift selection and subsequently pass.

\begin{figure*}[t]\centering
\subfloat[]{\includegraphics[width=.94\textwidth]{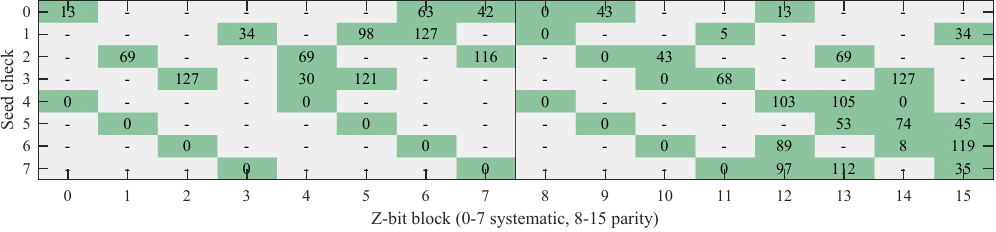}}\\[2pt]
\subfloat[]{\includegraphics[width=.94\textwidth]{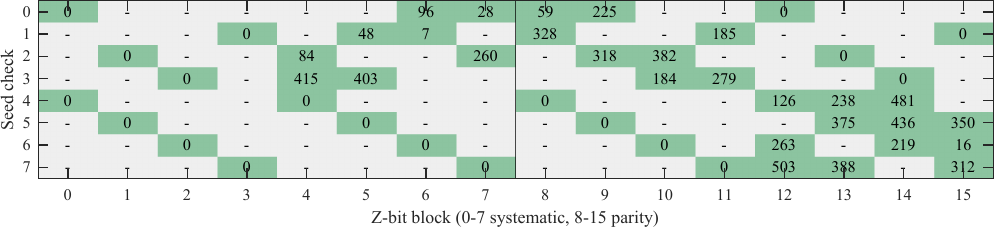}}\\[2pt]
\subfloat[]{\includegraphics[width=.94\textwidth]{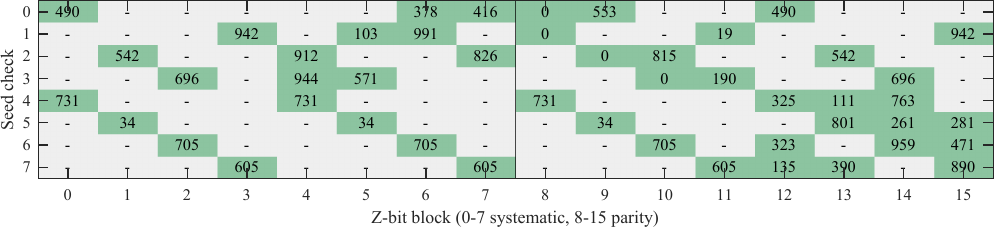}}
\caption{Equivalent rate-half QC checks: (a) code 5, $Z=128$, (b) codes 3/12, $Z=512$, and (c) code 10, $Z=1024$. Green cells identify circulant connections. A dash means no connection, and zero is an identity shift. The separator divides systematic and parity blocks. Z1024 rows are incidence-aligned to Z128 without rescaling shifts.}\label{fig:graphs}
\end{figure*}

\begin{figure*}[t]\centering
\subfloat[]{\includegraphics[width=.31\textwidth]{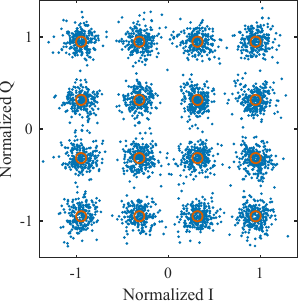}}
\subfloat[]{\includegraphics[width=.31\textwidth]{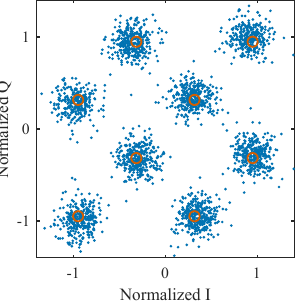}}
\subfloat[]{\includegraphics[width=.31\textwidth]{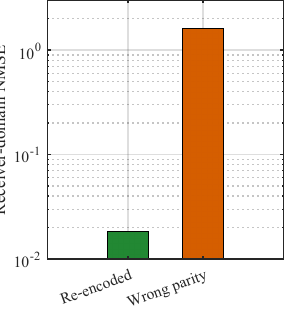}}
\caption{Frozen-encoder transfer: (a) code 39, (b) code 27, and (c) code-39 NMSE control. Blue points are equalized/calibrated observations; orange circles indicate remodulated alphabet points. Panel (b) uses the 2730 complete three-bit symbols of the code-27 field; the complete 2731-symbol NMSE is reported separately. Panel (c) compares code-39 re-encoding with a parity-flip control on a logarithmic scale. These figures do not provide additional independent target transmissions.}\label{fig:modulation}
\end{figure*}

The frozen $(8192,4096)$ encoder predicts two code-3 words in R13 and one QPSK code-12 word in frame R16, yielding 8192 and 4096 reserved parity bits, respectively, with zero disagreements. Check, parity-submatrix, and generator ranks are all 4096; check row weight is six. A separate code-12 refit reproduces the shifts but is not counted as a second independent validation. Fig.~\ref{fig:graphs} shows matching incidence across lifts; shift magnitudes do not scale linearly with $Z$.

Graph and repeated-shift relations precede Z512 numerical shift fitting. The remaining 448 positions test intra-word exclusion; separate code-3/code-12 words test transmission and modulation transfer. These validation scopes are distinct.

Shared incidence verifies an equivalent receiver graph pattern, not a single physical decoder or authenticated standard matrix. Numerical shifts and row gauge remain explicit in Fig.~\ref{fig:graphs}.

The single target-format field is R14 region 5 in Fig.~\ref{fig:atlasR14}. Only R14 contains the code-27 target field, beginning at slot 290, data ordinal $q=186$, carrier $k=-318$. The code-3/code-12 encoder is frozen. Searching all $8!$ labelings of the observed eight points yields map $[1,5,4,7,0,6,2,3]$ on sixteen-point grid indices $[0,2,5,7,8,10,13,15]$. The best hard labeling has 16 failed checks and the next exceeds 1200. The same three likelihood scales give one-iteration convergence after five hard-bit changes, with all 4096 checks and predicted parity bits agreeing.

The code-27 constellation plot uses the 2730 complete three-bit symbols, while the complete-field remodulation includes all 2731 symbols. For the last symbol, the 8192-bit candidate word leaves one unassigned position in the three-bit label. Setting that bit to one under the declared receiver polarity gives a full-field NMSE of 0.02843979 and a last-symbol NMSE of 0.01833607. This choice is consistent with the cross-format endpoint model, but does not authenticate the transmitter's spare-bit value. The observed alphabet is an eight-point subset rather than a confirmed standard 8QAM implementation.

Both target mappings are selected on their only available target field. Thus these are frozen-encoder transfer tests with same-field mapping selection, not independent same-format holdout experiments. The code-28 and code-37 core tests below separately verify the adopted eight-point labels and declared polarity against their own parity relations.

\subsection{Rate-5/8 QC-LDPC Recovery for Code 9}
R01 region 4 in Fig.~\ref{fig:atlasR01} begins at $(14,696,+196)$ and supplies 2048 BPSK symbols per word. The recovered parameters are $(N_c,K_c,Z)=(2048,1280,128)$, leaving 768 parity coordinates.

Code 9 uses 18 words from R01, R05, and R08 to recover a full input basis. Sparse-support projection and full-coordinate checks give $H$ of rank 768 and row weights 7--16. Thirteen reserved words in R12 and R15 predict 9984 parity bits without disagreement. All 31 words pass.

\subsection{Code 21 Recovery and Code 37 Shortening}
R00 region 4 in Fig.~\ref{fig:atlasR00} starts at $(14,696,+196)$ and uses QPSK. Code 21 has $(N_c,K_c,Z)=(8192,5120,512)$ and 3072 parity bits. The accepted quarter-phase labels map to two-bit integers $[3,1,0,2]$ in the receiver convention.

Code 21 uses 20 words from R00 and R04 and QC512 closure to span input rank 5120. Module-based recovery over $\GF[x]/(x^{512}+1)$ and sparse-support searches give check rank 3072 with row weights 7--18. The frozen model passes 264 reserved words with 811008 parity bits and zero disagreements. Input-basis rank 5120 is not a claim that the $5120\times3072$ matrix $P$ has that rank. Direct polynomial lifting from code 9 did not reproduce code 21; a shared graph is not asserted.

Their matching rates and fourfold lift ratio suggest a direct lift, but the tested polynomial transfer fails code-21 observations. Independent recovery and 264 reserved words support code 21 instead. The failure excludes that transfer, not every possible sharing transformation.

Code 37 is tested against this frozen code-21 encoder rather than fitted as an unrelated code. Its seven words occur in R02, R04, R11, and R14 (one, one, two, and three words). R02 region 6 in Fig.~\ref{fig:atlasR02} occupies $(293,52,-452)$ through $(295,774,+274)$ and contains 2731 eight-point symbols. Set the first 384 mother-code input bits to zero and place the 4736 observed systematic bits in mother coordinates 384--5119. The final 3072 observed bits then agree with the mother-code parity. This identifies a shortened $(7808,4736)$ core of rate $37/61$. The 384 observed bits between the systematic segment and the parity segment remain unexplained and are excluded from this relation.

At likelihood scales 0.01, 0.02, and 0.04, all seven words pass the shortened checks and re-encoding of all $7\times3072=21504$ parity coordinates. At scale 0.02, four hard decisions change in total. Remodulation uses only the 2601 symbols per word whose label bits are all predicted by the core. Its NMSE is 0.016956--0.026212. The remaining 130 symbols per word contain unexplained, mixed, or spare positions and are omitted. These results establish the shortened core, not a complete 8192-bit code-37 encoder or its overall information rate.

\subsection{Verified Systematic and Parity Resources in R02}
A field label does not yet expose the internal redundancy layout. With the code-5 and code-10 relations above frozen, their receiver-ordered words can be mapped back to the same slot/carrier grid. Code 5 uses BPSK, so each coded bit occupies one data modulation symbol: the first 1024 symbols of each 2048-symbol word are systematic coordinates and the following 1024 are predicted parity coordinates. Five words repeat that pattern after the first PDU header. Code 10 uses QPSK, so its first 8192 bits occupy 4096 symbols and its final 8192 parity bits another 4096. The system/parity transition can therefore lie inside an OFDM slot, just as an allocation boundary does.

Table~\ref{tab:parts} maps saved validated coding coordinates to inclusive slot/q/k endpoints. Code-5 parity prediction supports its split; code 10 retains its finite-completion qualification. ``Systematic'' denotes recovered encoder inputs, which can still include source scrambling and upper-layer fields.

\begin{table*}[t]\centering
\caption{Recovered Systematic/Parity Parts in R02}\label{tab:parts}
\begin{tabular}{lrrrr}\toprule
Part & First $(s,q,k)$ & Last $(s,q,k)$ & Symbols & Coded bits\\\midrule
C5 W00 system. & (2,228,-276) & (3,247,-257) & 1024 & 1024 \\
C5 W00 parity & (3,248,-256) & (4,267,-237) & 1024 & 1024 \\
C5 W01 system. & (4,268,-236) & (5,287,-217) & 1024 & 1024 \\
C5 W01 parity & (5,288,-216) & (6,307,-197) & 1024 & 1024 \\
C5 W02 system. & (6,308,-196) & (7,327,-177) & 1024 & 1024 \\
C5 W02 parity & (7,328,-176) & (8,347,-157) & 1024 & 1024 \\
C5 W03 system. & (8,348,-156) & (9,367,-137) & 1024 & 1024 \\
C5 W03 parity & (9,368,-136) & (10,387,-117) & 1024 & 1024 \\
C5 W04 system. & (10,388,-116) & (11,407,-97) & 1024 & 1024 \\
C5 W04 parity & (11,408,-96) & (12,427,-77) & 1024 & 1024 \\
C10 W0 system. & (12,884,+384) & (16,963,+463) & 4096 & 8192 \\
C10 W0 parity & (16,964,+464) & (21,39,-465) & 4096 & 8192 \\
\bottomrule\end{tabular}
\par\smallskip\footnotesize Each code-5 word contains 2048 BPSK symbols. The code-10 word contains 8192 QPSK symbols. All listed coordinates exclude pilots and DC. PDU header coding blocks, unknown high-order encoders, and reference-symbol padding are separately identified in Fig.~\ref{fig:atlasR02}.
\end{table*}

\begin{figure*}[t]\centering
\subfloat[]{\includegraphics[width=.94\textwidth]{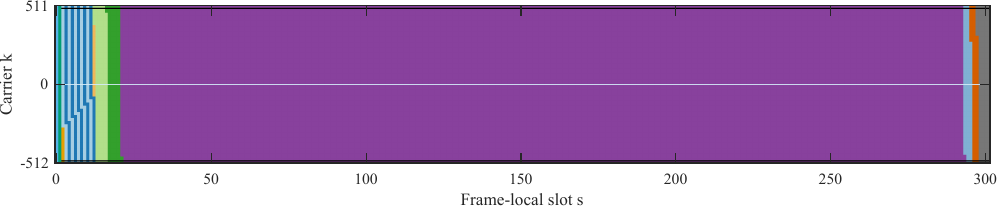}}\\[2pt]
\subfloat[]{\includegraphics[width=.94\textwidth]{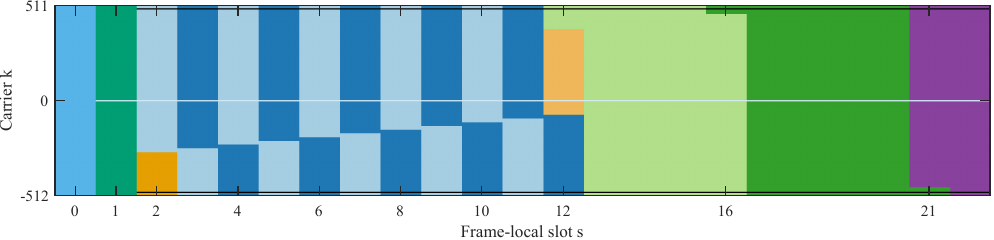}}\\[2pt]
\subfloat[]{\includegraphics[width=.94\textwidth]{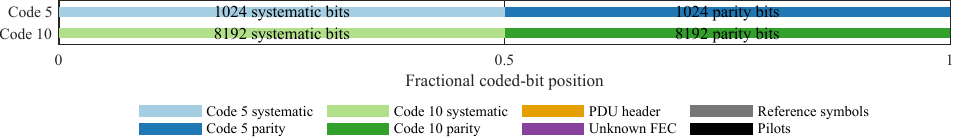}}
\caption{Validated coding overlay on the current R02 partition. The complete grid and early-slot enlargement preserve the corrected unknown-field boundaries. Light/dark blue and green mark systematic/parity coordinates for codes 5 and 10; the bottom panel shows their bit layouts. Unknown-field colors denote allocations only. Code-10 identification remains conditional on its declared finite completion family.}\label{fig:parts}
\end{figure*}

\begin{figure*}[p]\centering
\includegraphics[width=\textwidth]{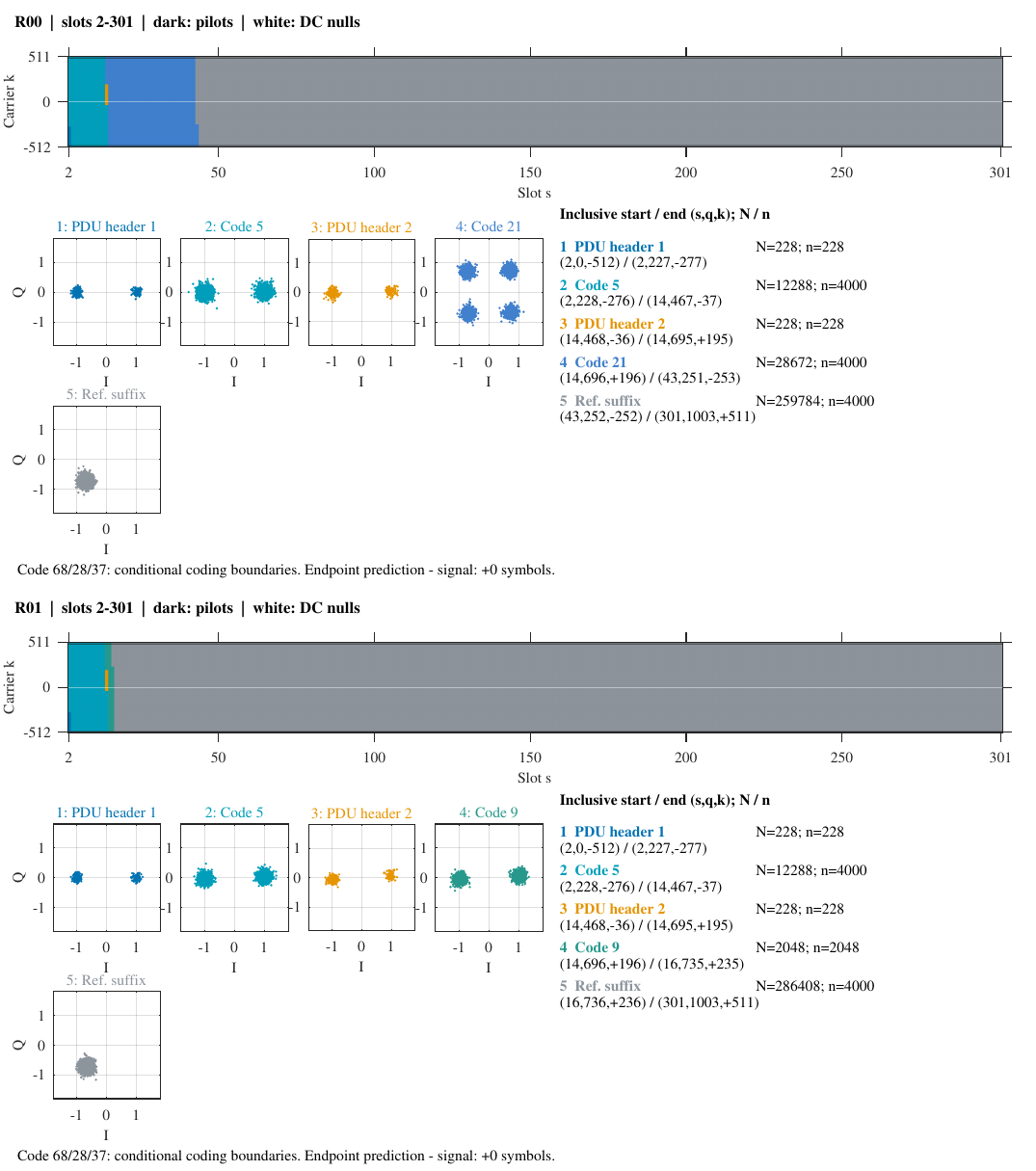}
\caption{Complete regions and reference-removed constellations for R00 (upper) and R01 (lower). Endpoints are inclusive. The tables give non-pilot symbol counts $N$ and displayed counts $n$.}\label{fig:atlasR00}\label{fig:atlasR01}
\end{figure*}

Fig.~\ref{fig:parts} uses light/dark blue for code-5 system/parity and light/dark green for code 10. The bottom panel is fractional coded-bit position; the carrier maps show simultaneous OFDM resources. The panels display only the code-5 and code-10 redundancy partitions. The shortened-core relations of codes 28 and 37 are evaluated separately below, and code 68 remains unresolved.

This split enables a stronger experiment than demodulating and plotting a whole word. Keep the observed systematic coordinates fixed, predict parity through the frozen matrix, restore the declared bit labels and PN phase, and compare only that prediction with received parity or with the full remodulated field. A model that simply repeats the received hard decisions cannot pass this test by construction. The separate parity-flip control preserves the alphabet while violating the learned relation, isolating the redundancy contribution to the signal-domain explanation.

\subsection{Shortened LDPC Core Recovery for Code 28}
\begin{figure*}[p]\centering
\includegraphics[width=\textwidth]{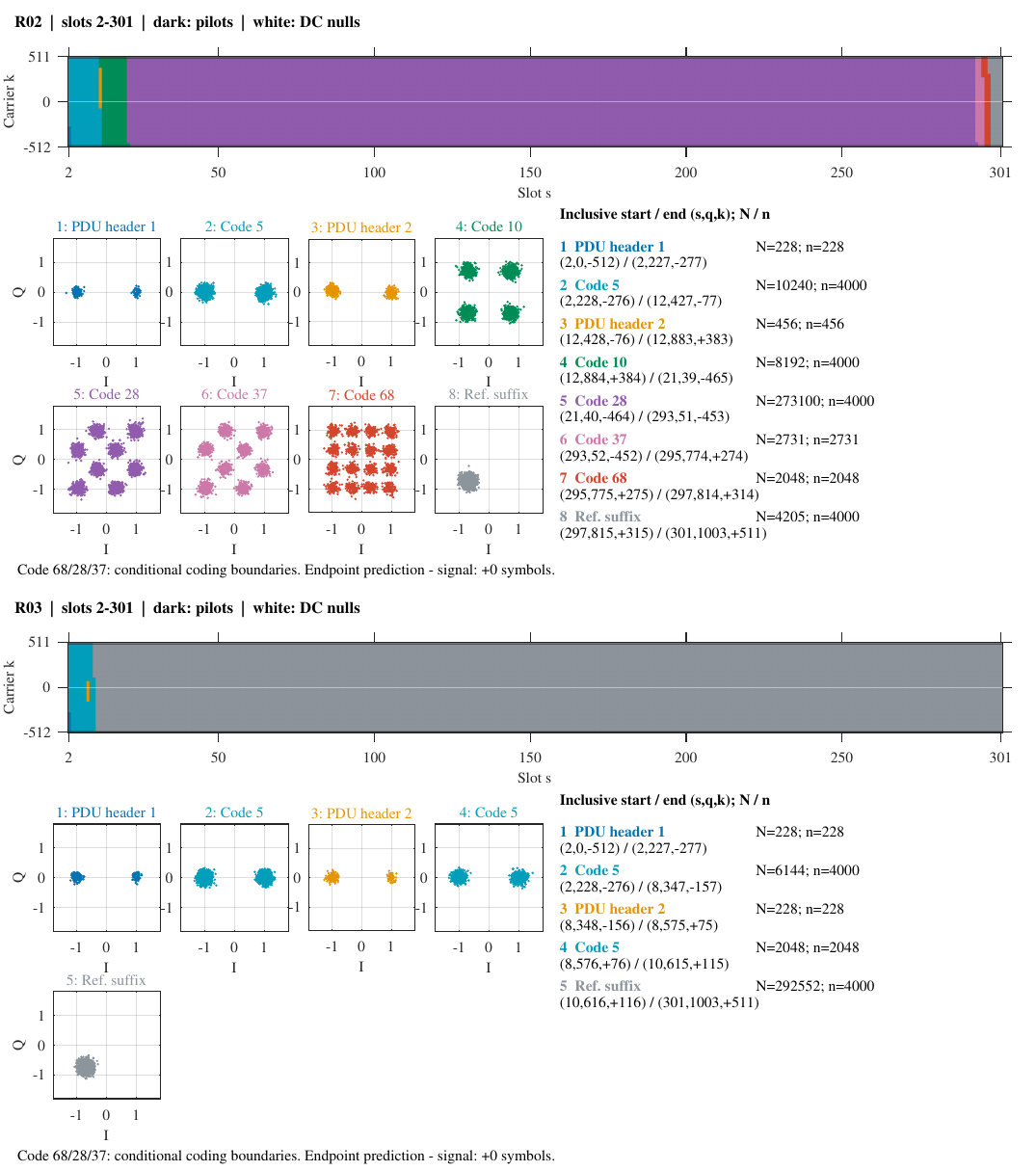}
\caption{Complete regions and reference-removed constellations for R02 (upper) and R03 (lower). Endpoints are inclusive. The tables give non-pilot symbol counts $N$ and displayed counts $n$.}\label{fig:atlasR02}\label{fig:atlasR03}
\end{figure*}

\begin{figure*}[p]\centering
\includegraphics[width=\textwidth]{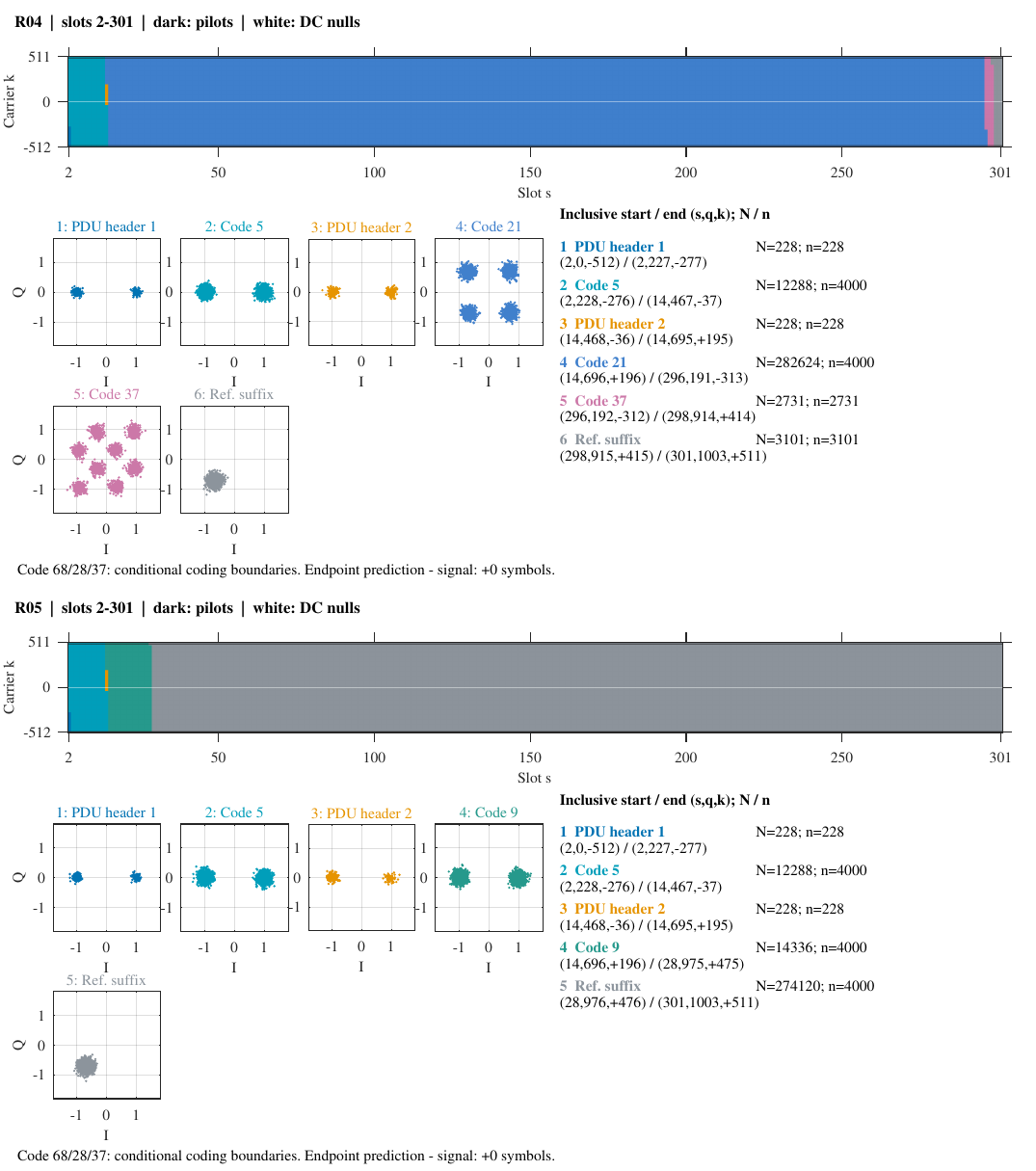}
\caption{Complete regions and reference-removed constellations for R04 (upper) and R05 (lower). Endpoints are inclusive. The tables give non-pilot symbol counts $N$ and displayed counts $n$.}\label{fig:atlasR04}\label{fig:atlasR05}
\end{figure*}

R02 region 5 in Fig.~\ref{fig:atlasR02}, from $(21,40,-464)$ to $(293,51,-453)$, contains 50 words of 5462 eight-point symbols. The corresponding fields in R06, R09, and R13 contain 50, 41, and 48 words. Thus 189 words are available. With the accepted bit labels and a global polarity convention, the QC1024 equations from the 100 R02/R06 words determine a mother encoder with $(N_c,K_c)=(16384,9216)$. The recovered generator has nine systematic cyclic blocks and seven parity cyclic blocks. Its sparse dual supplies 7168 independent checks.

To test shortening, fix the first 256 mother-code input bits to zero. Map the first 8960 observed bits to mother systematic coordinates 256--9215 and use the final 7168 observed bits as parity. Removing the fixed coordinates gives a $(16128,8960)$ core of rate $5/9$, with a rank-7168 check matrix. The intervening 256 observed bits at received positions 8960--9215 are not explained by this construction. They are not substituted for the fixed mother-code coordinates. The frozen encoder predicts all $89\times7168=637952$ parity bits of the R09/R13 words with zero disagreement.

Soft decoding at scales 0.01, 0.02, and 0.04 passes all 189 words. At scale 0.02, 57 words require 80 hard-bit changes in total. Re-encoding agrees at all $189\times7168=1354752$ core parity positions. Core-only remodulation evaluates 5375 of the 5462 symbols per word and gives NMSE 0.016995--0.022086. A parity-flip control gives 1.5264--1.6310 on the same set. The 87 remaining symbols contain unexplained or mixed positions. Neither their values nor a complete 16384-bit encoder are recovered by the core test.

US12074683B1 describes appending pseudorandom bits when LDPC output is too short to fill the allocated OFDM resources~\cite{mccormick2024ofdm}. This provides a possible role for the extra 256 and 384 bits of codes 28 and 37. Their generation rule has not been recovered. Moreover, in the receiver ordering used here these bits lie between the identified systematic and parity segments, whereas the patent describes appending to the LDPC output. No intervening permutation or alternative serialization has been established to reconcile those positions. We therefore retain pseudorandom filling as a candidate explanation, not a demonstrated identification of these bits. It is also distinct from the one or two unused label positions in a final three-bit modulation symbol and from the frame-level reference-symbol padding.

\subsection{Unresolved Channel Coding of Code 68}
For code 68, the R02 allocation begins at $(295,775,+275)$. Using the allocation boundaries supported by the length checks, seven candidate words tested with known Gray/code combinations fail: the code-3 checks leave 1989--2068 of 4096 equations nonzero, and the code-21 checks leave 1517--1564 of 3072 nonzero. A declared arbitrary sixteen-point/four-bit mapping search does not produce a valid bijection. These results reject the tested mapping/code hypotheses at the supported boundaries, not all code-68 encoders.

The present evidence does not identify a code-68 generator, parity-check matrix, information dimension, or rate. Its field position and sixteen-point observations remain supported separately from its unresolved coding. The format label is retained under the control-field convention of Section IV.

Table~\ref{tab:codes} summarizes the recovered parameters and the parity predictions reported for each coding family. The listed counts refer to the stated validation words or target fields, not independent bit trials.

\begin{table*}[t]\centering
\caption{Recovered Code Representations and Validation Scope}\label{tab:codes}
\begin{tabular}{rrrrllr}\toprule
Code & $N_c$ & $K_c$ & $Z$ & Alphabet & Reserved/target parity bits & Disagreements\\\midrule
5 & 2048 & 1024 & 128 & BPSK & 78848; 77 reserved words & 0\\
9 & 2048 & 1280 & 128 & BPSK & 9984; 13 reserved words & 0\\
21 & 8192 & 5120 & 512 & QPSK & 811008; 264 reserved words & 0\\
10 & 16384 & 8192 & 1024 & QPSK & 8192; 1 word excluded from fitting & 0\\
3 & 8192 & 4096 & 512 & BPSK & 8192; 2 reserved words & 0\\
12 & 8192 & 4096 & 512 & QPSK & 4096; 1 reserved cross-format word & 0\\
39 & 16384 & 8192 & 1024 & 16QAM & 8192 within the only target field & 0\\
27 & 8192 & 4096 & 512 & Eight-point subset & 4096 within the only target field & 0\\
28 core & 16128 & 8960 & 1024 & Eight-point subset & 637952; 89 reserved words & 0\\
37 core & 7808 & 4736 & 512 & Eight-point subset & 21504; 7 target words & 0\\
\bottomrule\end{tabular}
\par\smallskip\footnotesize Codes 10 and 3/12 rely on their stated completion/graph families. Codes 39/27 have no same-format held-out transmission; their mappings were selected on the target field. Codes 28/37 have the shortened-core results stated above, not complete-field encoder recovery. Code 68 remains unresolved.
\end{table*}

All 18 records belong to one continuous capture. Some observations were inspected during model selection, so exclusion from the final generator fit does not guarantee previously unseen validation data~\cite{cawley2010}. In particular, code-10 R13 supplies consistency evidence, codes 39 and 27 each have only one target-format field, and codes 28 and 37 are validated only over their identified cores. Zero parity disagreement is measured against recovered receiver bits rather than transmitted truth and does not establish channel BER or protocol identity. No new capture acquired after all models were fixed is included.

\section{Comparison With Published Modulation and Coding Configurations}
Table~\ref{tab:patentcomparison} compares the formats extracted from slot 2 onward with the example MCS table in Fig. 4 of US11792832B1~\cite{iyer2023amc}. That patent states that the displayed rows are only a subset and that their values are examples. A missing row therefore means no direct comparison is available from that published subset, not that the format is absent from the system. The numeric correspondence is evaluated after the receiver labels and coding relations have been established. It is not used to redefine the control-field bit split in Section IV.

\begin{table*}[t]
\centering\small
\caption{Observed Formats and the Published Example MCS Table}\label{tab:patentcomparison}
\begin{tabular}{p{.065\textwidth}p{.345\textwidth}p{.28\textwidth}p{.21\textwidth}}\toprule
Format & Verified observation & Patent example & Correspondence\\\midrule
PDU header & BPSK, BCC rate $1/3$, 32 retained bits $+6$ tails, 114 output bits & 0: BCC--BPSK, $N=114$, $K=32$ & Lengths and encoder type agree\\
10 & QPSK, $(16384,8192)$, rate $1/2$, 8192 symbols & 10: same parameters & Parameter match, conditional completion\\
12 & QPSK, $(8192,4096)$, rate $1/2$, 4096 symbols & 12: same parameters & Parameter match\\
39 & 16QAM, $(16384,8192)$, rate $1/2$, 4096 symbols & 39: same parameters & Parameter match and frozen-encoder transfer\\
27 & Eight points, $(8192,4096)$, rate $1/2$, 2731 symbols & 27: 8PSK, same lengths/rate/symbol count & Coding and resource match, alphabet differs\\
28 & Eight points, 5462 symbols, $(16128,8960)$ core $+256$ unexplained bits & 28: 8PSK, $(16384,8960)$, rate $35/64$, 5462 symbols & Several lengths agree, full code unconfirmed\\
3 & BPSK, $(8192,4096)$, rate $1/2$ & No matching row in published subset & No direct correspondence\\
5 & BPSK, $(2048,1024)$, rate $1/2$ & No matching row in published subset & No direct correspondence\\
9 & BPSK, $(2048,1280)$, rate $5/8$ & No matching row in published subset & No direct correspondence\\
21 & QPSK, $(8192,5120)$, rate $5/8$ & No matching row in published subset & No direct correspondence\\
37 & Eight points, $(7808,4736)$ core $+384$ unexplained bits & Index 37 not shown & Complete scheme unconfirmed\\
68 & Sixteen-point observations, encoder/rate unknown & Index 4 not shown & Unresolved\\
\bottomrule\end{tabular}
\end{table*}

For the PDU header row, the convolutional encoder produces three output bits for each of the 38 input clocks, including six termination clocks. The 32 retained bits include the header CRC contribution in a final block. Thus the rate-$1/3$ encoder convention and the patent's $K=32$ do not imply an effective retained-bit ratio of $1/3$ for 114 transmitted bits.

Codes 27 and 28 occupy an eight-point subset of a sixteen-point grid after the declared PN removal, with unequal point magnitudes. The corresponding example rows name 8PSK. Numerical length and rate agreement therefore cannot establish constellation identity. Code 28's 8960 recovered information coordinates and 16384 nominal transmitted positions give the numerical ratio $35/64$, but its verified core rate is $8960/16128=5/9$. Neither ratio explains the extra 256 bits. Code 37 similarly has a verified core rate $4736/7808=37/61$, not a demonstrated complete-field rate.

The downlink patent also lists 8QAM among possible modulation formats~\cite{mccormick2024ofdm}, but that general statement does not identify these particular eight-point mappings. The patent comparisons establish compatible parameters or candidate mechanisms, not equality of parity-check matrices or authenticated transmitter-standard identities. Equivalent basis changes, receiver labeling, and phase conventions can change the displayed representation without changing its observable code space. Additional acquisitions evaluated after all models are fixed remain necessary to assess generalization.

\section{Conclusion}
The continuous capture supports terminated rate-$1/3$ control blocks, a header CRC, and a conditional allocation layout checked against received-symbol boundaries. Full-rank sparse QC representations, reserved-word parity predictions, and frozen-encoder transfer establish selected receiver-equivalent code relations. Codes 28 and 37 further admit shortened LDPC cores, while their extra 256 and 384 bits remain unexplained. Published patent configurations agree with several observed lengths and rates but do not establish complete encoder identity, and the eight-point alphabets differ from the example 8PSK labels. Code 68 remains unresolved. New acquisitions evaluated after all models are fixed are needed to test generalization beyond this capture.

\appendices
\section{Complete Frame Regions and Reference-Removed Constellations}
\label{app:regions}
All panels use measured reference-removed symbols with one frame-wide gain normalization determined from the first 228 PDU header resources. No region-specific phase fit or hard-decision replacement is used. Each frame covers slots 2--301 and all 301200 non-pilot resources. Region numbers connect the map, boundary table, and constellation panels. Pilots are dark and DC nulls are white. Code-68/28/37 boundaries retain their conditional status. R07 and R16 retain descriptor-minus-signal endpoint differences of one and two symbols. Each constellation uniformly displays at most 4000 actual observations. The complete arrays remain in the original saved results.

\clearpage
\clearpage
\clearpage
\clearpage
\begin{figure*}[p]\centering
\includegraphics[width=\textwidth]{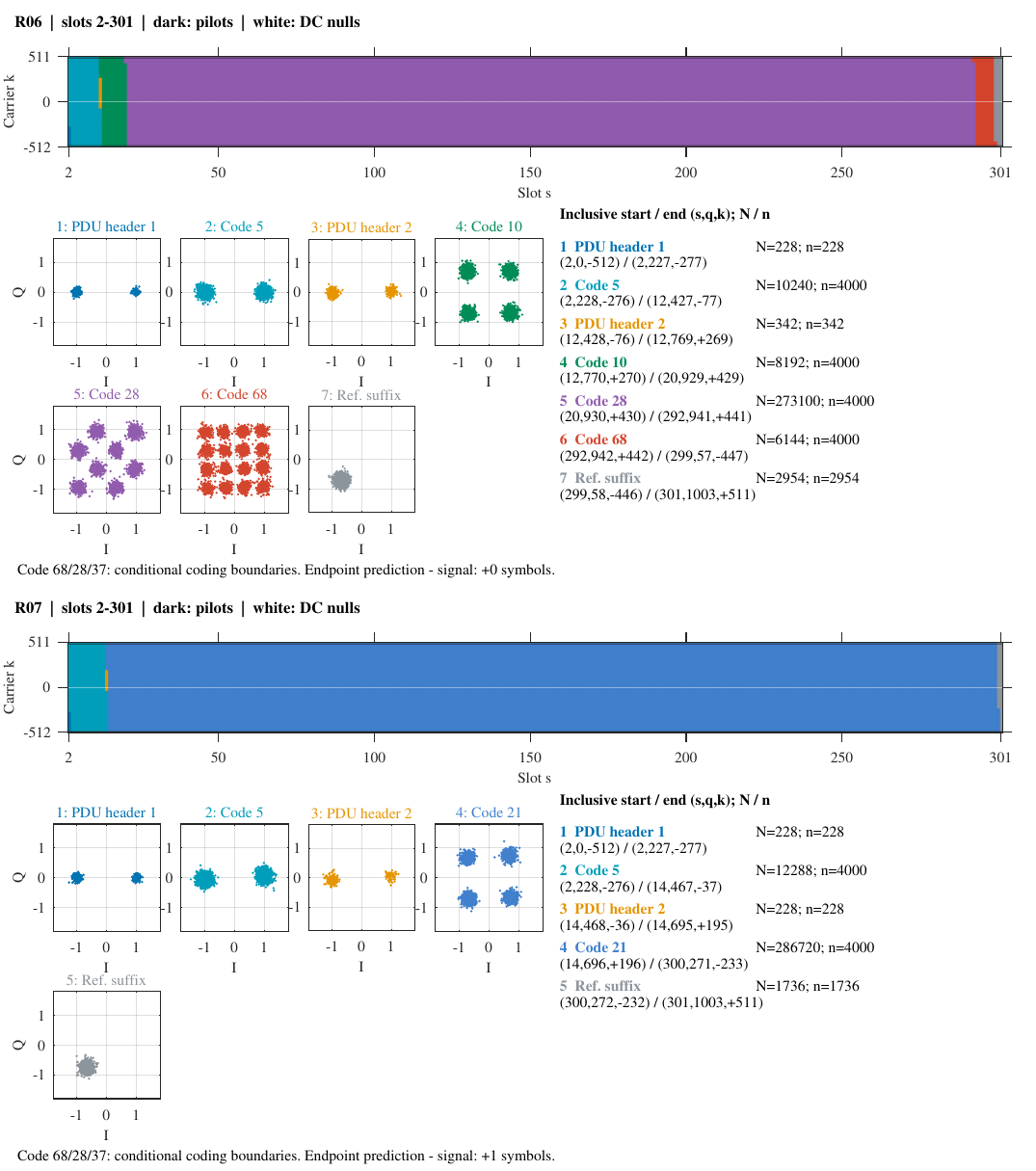}
\caption{Complete regions and reference-removed constellations for R06 (upper) and R07 (lower). Endpoints are inclusive. The tables give non-pilot symbol counts $N$ and displayed counts $n$.}\label{fig:atlasR06}\label{fig:atlasR07}
\end{figure*}
\clearpage
\begin{figure*}[p]\centering
\includegraphics[width=\textwidth]{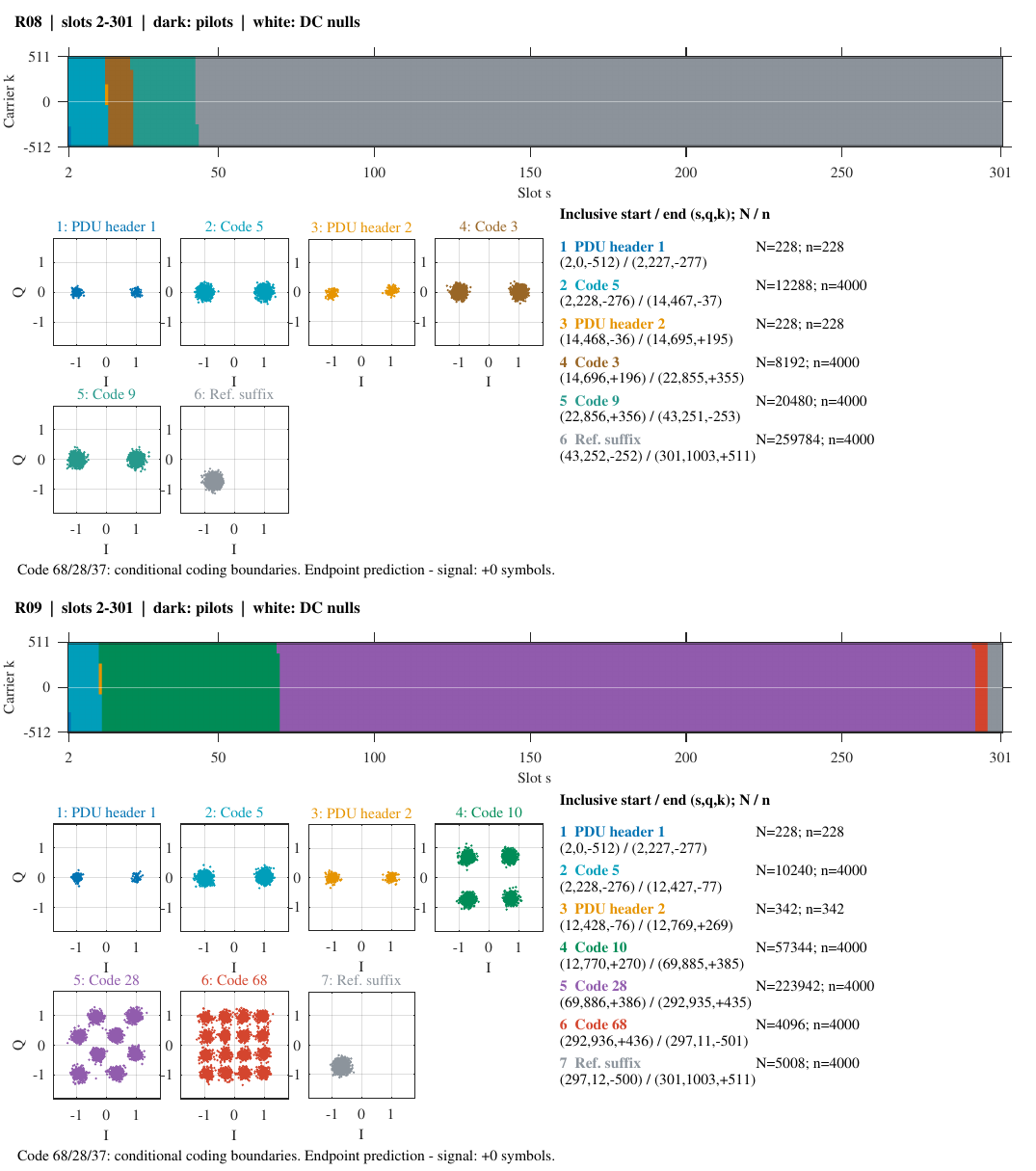}
\caption{Complete regions and reference-removed constellations for R08 (upper) and R09 (lower). Endpoints are inclusive. The tables give non-pilot symbol counts $N$ and displayed counts $n$.}\label{fig:atlasR08}\label{fig:atlasR09}
\end{figure*}
\clearpage
\begin{figure*}[p]\centering
\includegraphics[width=\textwidth]{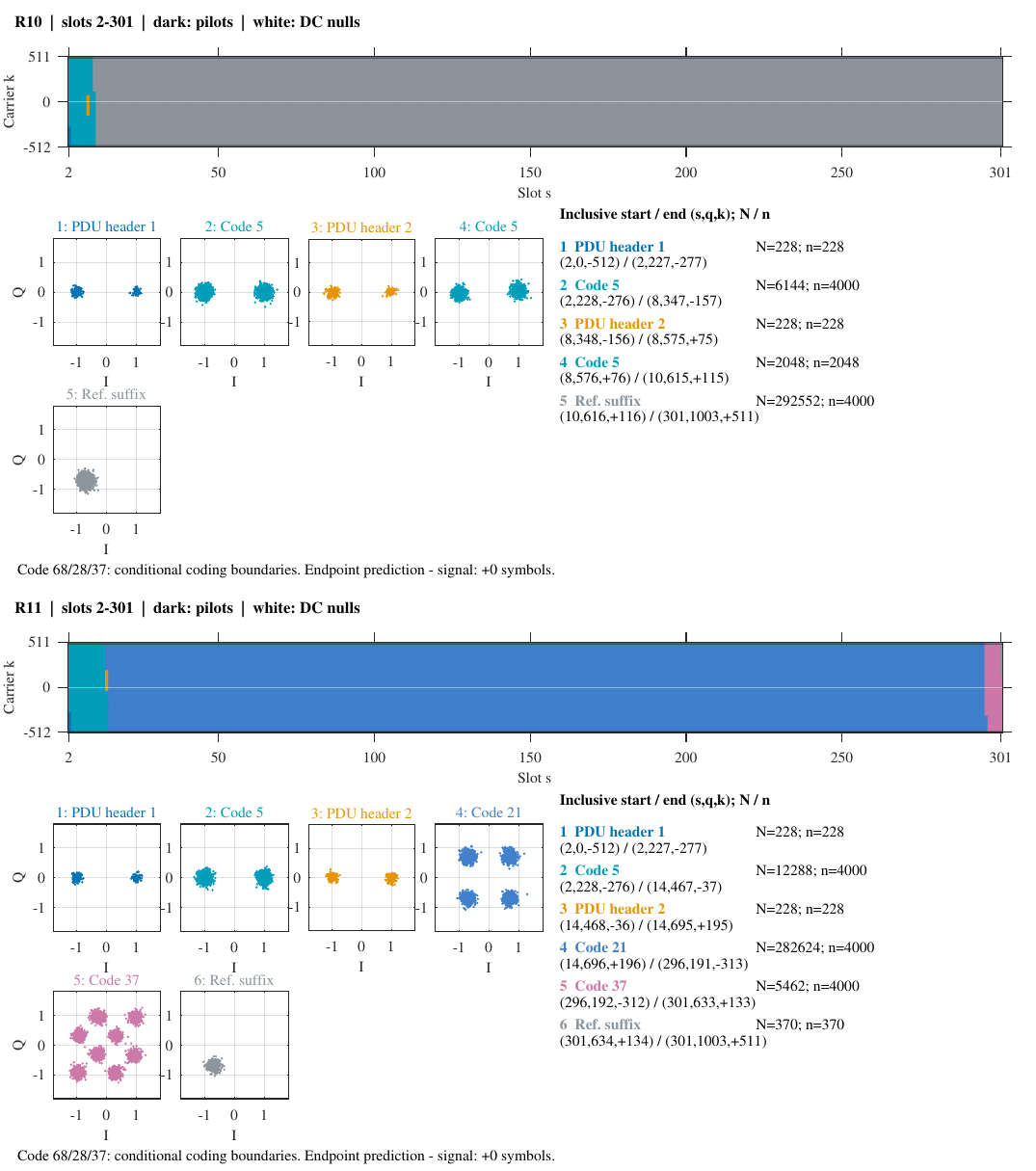}
\caption{Complete regions and reference-removed constellations for R10 (upper) and R11 (lower). Endpoints are inclusive. The tables give non-pilot symbol counts $N$ and displayed counts $n$.}\label{fig:atlasR10}\label{fig:atlasR11}
\end{figure*}
\clearpage
\begin{figure*}[p]\centering
\includegraphics[width=\textwidth]{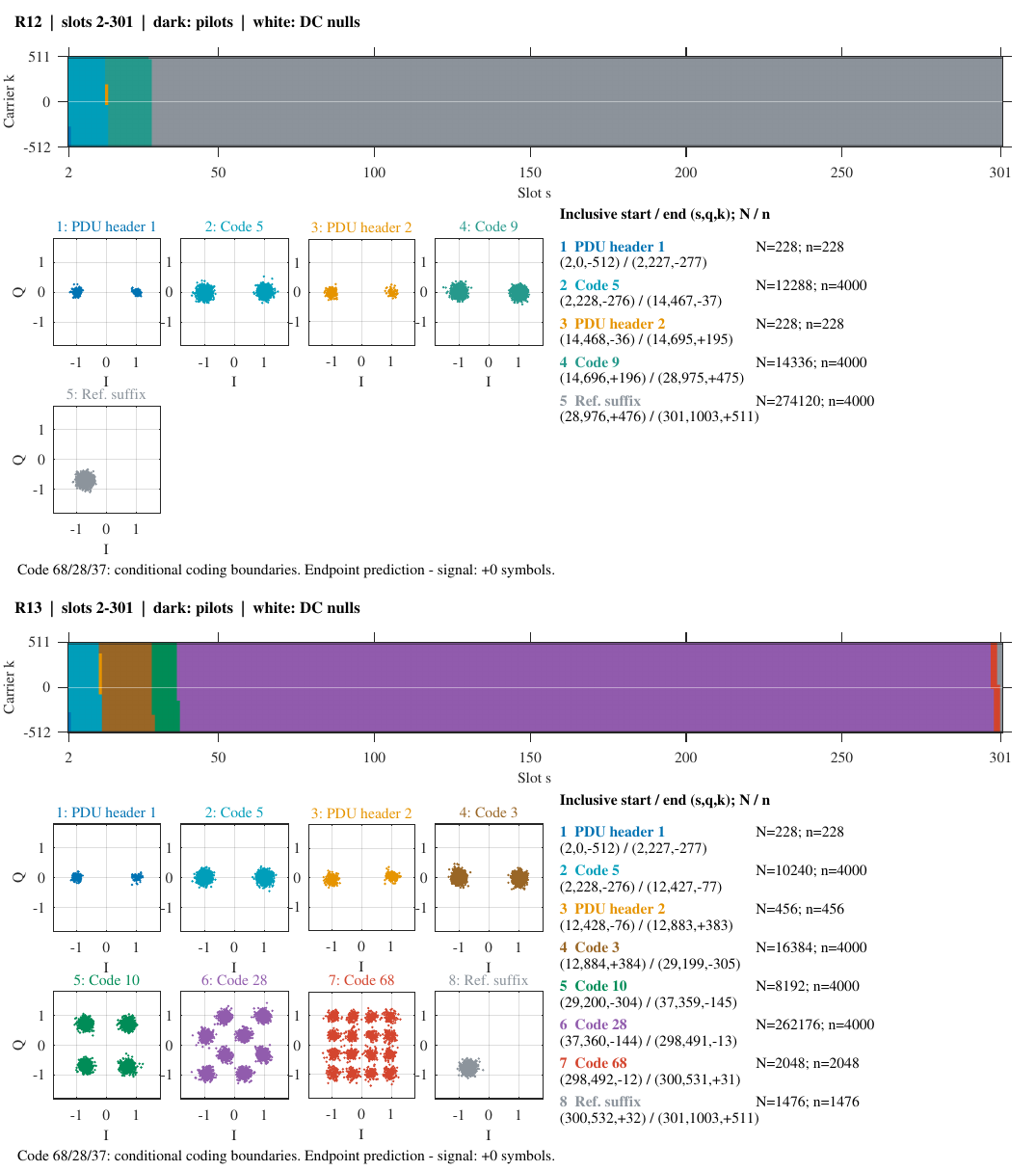}
\caption{Complete regions and reference-removed constellations for R12 (upper) and R13 (lower). Endpoints are inclusive. The tables give non-pilot symbol counts $N$ and displayed counts $n$.}\label{fig:atlasR12}\label{fig:atlasR13}
\end{figure*}
\clearpage
\begin{figure*}[p]\centering
\includegraphics[width=\textwidth]{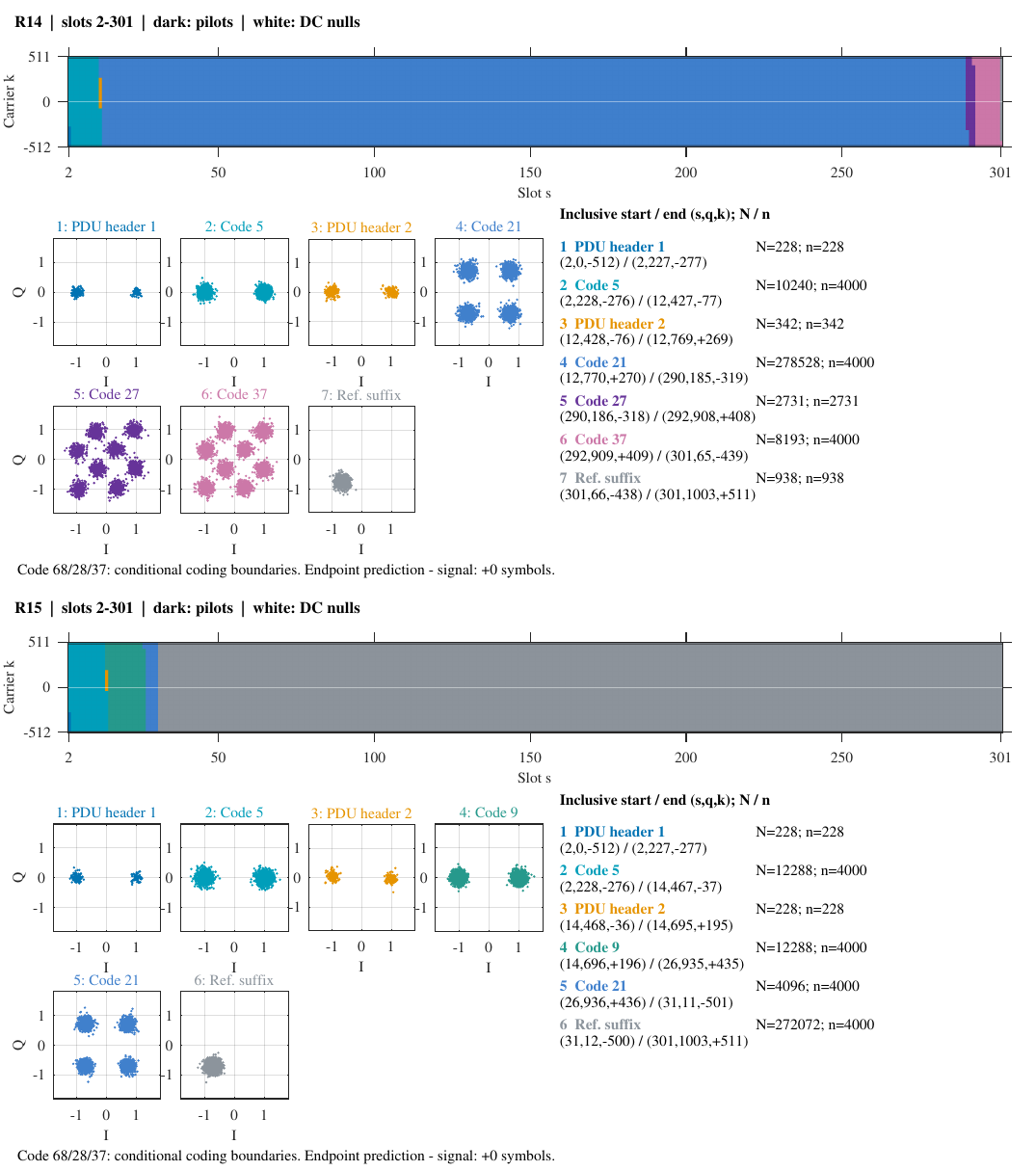}
\caption{Complete regions and reference-removed constellations for R14 (upper) and R15 (lower). Endpoints are inclusive. The tables give non-pilot symbol counts $N$ and displayed counts $n$.}\label{fig:atlasR14}\label{fig:atlasR15}
\end{figure*}
\clearpage
\begin{figure*}[p]\centering
\includegraphics[width=\textwidth]{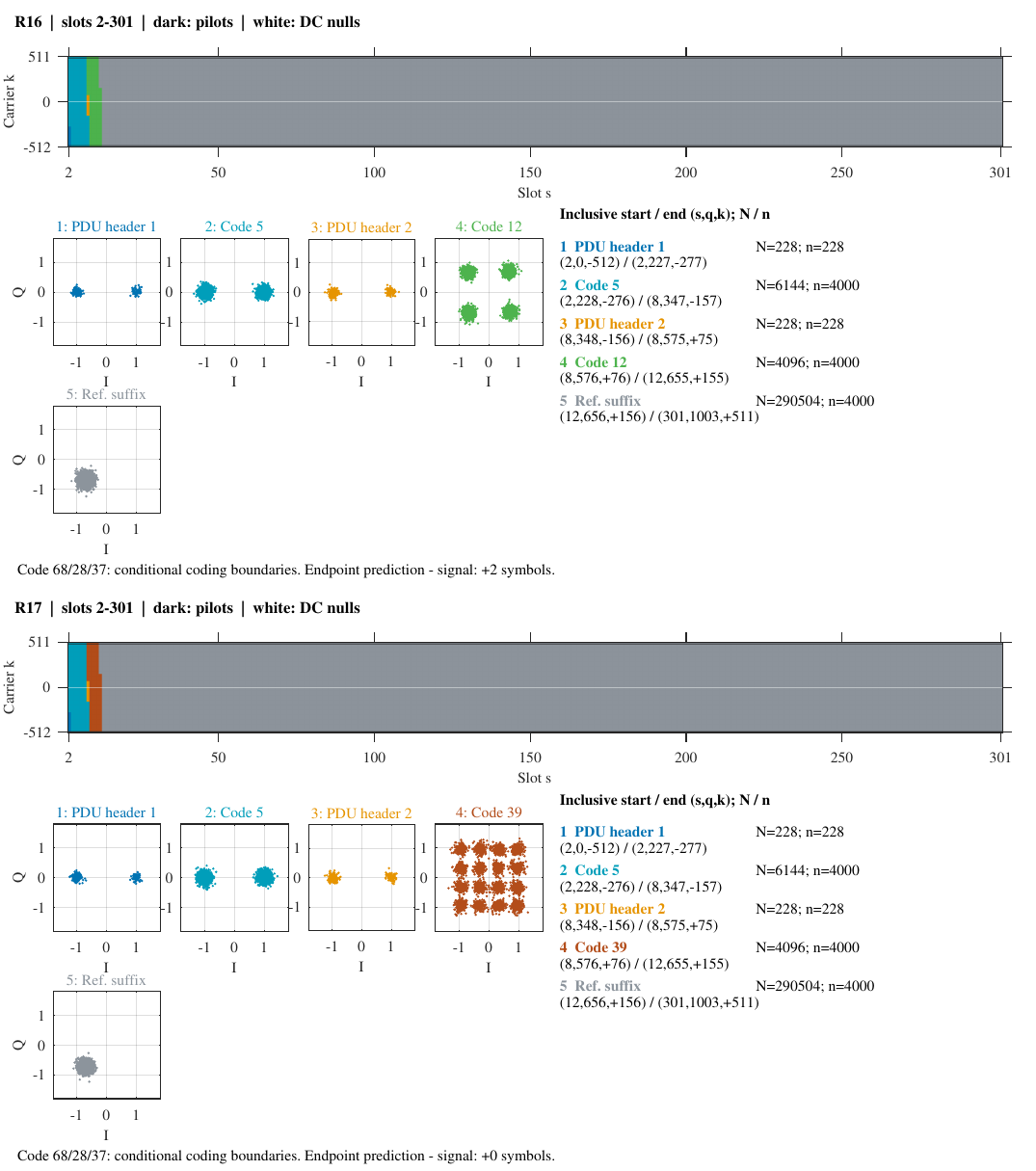}
\caption{Complete regions and reference-removed constellations for R16 (upper) and R17 (lower). Endpoints are inclusive. The tables give non-pilot symbol counts $N$ and displayed counts $n$.}\label{fig:atlasR16}\label{fig:atlasR17}
\end{figure*}
\clearpage

\twocolumn
\bibliographystyle{IEEEtran}
\bibliography{references}
\end{document}